\documentclass[reprint,amsmath,amssymb,aps,nofootinbib,superscriptaddress]{revtex4-2}

\usepackage{graphicx}
\usepackage{dcolumn}
\usepackage{bm}
\usepackage{numprint}
\usepackage[compat=1.1.0]{tikz-feynman}
\usepackage{tikz}
\usepackage{svg}
\usepackage{subfigure}
\usepackage{eqnarray}
\usepackage{hyperref}
\usepackage[mathlines]{lineno}
\usepackage{cancel}
\usepackage{amsmath}
\usepackage{booktabs} 

\usepackage{overpic}
\usepackage{diagbox}
\usepackage{makecell}
\usepackage{multirow}

\tikzfeynmanset{warn luatex=false}
\newcommand{\nflows}{$\nu^2$-flows }
\DeclareMathOperator*{\concat}{%
\mathchoice%
    {\big\Vert}%
    {\big\Vert}%
    {\Vert}%
    {\Vert}%
}

\begin{document}

\title{Comprehensive reconstruction of collider events with hypergraph representation learning and graph-conditioned diffusion}

\author{Lining Mao}
 \affiliation{School of Physics and Astronomy, Shanghai Jiao Tong University, No.800 Dong Chuan Road, Shanghai 200240, PRC}
\author{Yvonne Peters}
\author{Ethan Simpson}
 \email{ethan.simpson@manchester.ac.uk}
\author{Zihan Zhang}
 \email{zihan.zhang@manchester.ac.uk}
 \affiliation{Department of Physics and Astronomy, University of Manchester,\\Oxford Road, Manchester M13 9PL, UK}

\date{\today}

\begin{abstract}
In particle collider experiments, event reconstruction is the task of inferring the kinematics of short-lived particles produced in the hard scatter from the stable final states recorded by detectors. 
We decompose event reconstruction into two primary tasks: assigning measured jets and charged leptons to parent particles, and predicting unmeasured neutrino kinematics. 
We present VyPER, a novel geometric learning framework that represents collider events as hypergraphs with a physics-inspired topology. 
VyPER combines the supervised classification of hyperedges for particle assignment with a diffusion model for predicting neutrino kinematics, 
leveraging a joint loss function to optimize both reconstruction tasks within a unified framework.
We showcase VyPER across several proton–proton collision processes, comparing its performance to existing analytical and machine-learning-based reconstruction techniques. 
In doing so, we demonstrate that accurate event reconstruction is achievable across a diverse range of Standard Model physics processes, opening new avenues for precision measurements in the Higgs boson, electroweak, and top-quark sectors.
\end{abstract}

\maketitle

\section{\label{sec:introduction}Introduction}
Experiments at the Large Hadron Collider (LHC) at CERN~\cite{evans2008lhc} collide protons at high energies to measure properties of Standard Model (SM) particles and search for signs of physics beyond the SM.
The kinematics of the heavy, short-lived particles produced in the initial high-energy interaction — the hard scatter — serve as a critical probe into underlying SM dynamics and a vital tool for uncovering subtle deviations from known physics.
Observables constructed from the kinematics of these short-lived particles are used to probe the spins and polarizations of heavy fermions~\cite{bernreuther2015set} and bosons \cite{ballestrero2020different}; 
such measurements provide insight into the CP properties of Higgs boson couplings~\cite{Anderson:2013afp}, the mechanism of electroweak symmetry breaking~\cite{Lee:1977eg}, and the nature of quantum correlations between SM constituents~\cite{afik2021entanglement,barr2022testing}. 
Unfolded differential cross-sections of top quark kinematics provide direct comparison to fixed-order predictions of the SM \cite{aaboud2016measurement, sirunyan2017measurement}.
Such observables even enhance the separation of signal processes from complex backgrounds~\cite{sirunyan2018evidence, aad2022measurement}, and may be beneficial to experimental calibration procedures~\cite{atlas2025calibration}.

Heavy states produced in the hard scatter of proton-proton collisions decay almost instantly, preventing direct measurement of their properties. 
General purpose detectors~\cite{ATLAS:2008xda,CMS:2008xda} instead record the signals left by their stable or long-lived decay products.
These signals are reconstructed into calibrated final-state objects like jets and charged leptons for physics analysis. 
Measuring observables constructed from the kinematics of the parent particles can provide greater sensitivity and physical interpretability than measurements restricted to only the kinematics of the recorded final states.
Thus, the utility and physics reach of parent particle observables motivate developing accurate ``event reconstruction'' techniques.

With the advent of powerful machine learning (ML) architectures, substantial performance gains have been realized in event reconstruction.
Such models have generally focused on specific sub-tasks within event reconstruction, limiting their applicability to specific physics processes.
The utility of event reconstruction across the gamut of SM measurements remains under-explored.
In this paper, we showcase the potential of event reconstruction in top-quark, Higgs boson and electroweak physics processes, and present VyPER, a dedicated ML model designed to provide comprehensive event reconstruction in a variety of proton-proton scattering processes.

\section{\label{sec:state_of_the_art}Problem Specification and State of the Art}
The goal of event reconstruction is to estimate the four-momenta of the parent particles whose decays led to the observed final state. 
Different approaches and challenges exist depending on the scattering process in question and the decay modes of the particles we wish to reconstruct. 
In this section, we discuss how event reconstruction can be broken down into two separate tasks: final-state assignment and neutrino prediction. 
We highlight current methods in the literature built to perform each task.

\subsection{Jet and charged lepton assignment}
Jets and charged leptons produced in the decays of heavy, short-lived particles can be recorded with high-efficiency over almost the entire fiducial phase space at general-purpose detectors like ATLAS~\cite{ATLAS:2008xda} and CMS~\cite{CMS:2008xda}. 
The \textit{assignment} task associates these recorded decay products to their parent particle,
 partitioning the set of all measured final states into groups corresponding to the specific decay products of each parent\footnote{Decay products produced outside the detector's fiducial acceptance are not recorded, which can lead to incomplete associations.}.
 Summing the four-momenta of the final states in each partition then defines the parent's kinematics.

Traditionally solved through $\chi^2$--minimization techniques~\cite{chi2minimization2017}, machine learning approaches to the assignment problem have become a hot topic in high-energy physics studies and represent the state-of-the-art.
The majority of ML-based event reconstruction models use supervised classification to assign jets and charged leptons to parent particles.
This requires recorded final-state objects to be uniquely labeled, typically through angular matching of the detector-level objects to the decay products of the parent particles within the simulation truth record.
A variety of transformer~\cite{spanet2021,spanet2022,spanet2024,lee2024zero,heo2025improving}, message-passing graph-based \cite{hyper2024,topograph2023} and joint transformer-graph~\cite{soybelman2026topology,hermansen2026pairton} architectures are available in the literature.
Relevant to this study are the HyPER model~\cite{hyper2024} upon which the VyPER architecture is built, and the SPANet model~\cite{spanet2021,spanet2022,spanet2024} which has long served as the benchmark in assignment studies, and provides comparison to VyPER in Sections~\ref{sec:tt2L} and~\ref{sec:ttW}.
A short review of the individual approaches each takes to event reconstruction is given in Appendix~\ref{sec:appendix_assing}.

Alternative unsupervised methods formulate a learning task that does not rely on simulation truth labels. 
Approaches include training autoencoders directly on experimental data \cite{badea2024data}, and reformulating assignment as a sequential Markov decision process solved via reinforcement learning using a transformer-based agent~\cite{Dillon:2025dxr}.
Such approaches are not considered in this paper.

\subsection{Neutrino reconstruction}
Final states involving neutrinos pose a different challenge, as collider detectors~\cite{ATLAS:2008xda,CMS:2008xda} are not designed to detect these weakly interacting particles. 
Through conservation of momentum in the transverse plane and judicious measurement of all detector activity, the missing transverse momentum ($E_T^{\text{miss}}$ or ``MET'') quantity can be built.
This serves as a proxy for unmeasured neutrino activity in the transverse plane in physics processes where no other unmeasurable final states are produced.

The second task in event reconstruction is to infer neutrino kinematics using the visible final states and missing transverse momentum.
The task is formidable, since it represents an inverse problem which is frequently ill-posed: the measured final-state kinematics provide only incomplete constraints on the unobserved neutrino degrees of freedom, and the underlying distribution of neutrino kinematics can be highly non-Gaussian and multi-modal.

Traditional solution methods introduce physical assumptions — such as assuming collinearity of the decay products or imposing known mass constraints on intermediate resonances — to construct a closed system of kinematic equations~\cite{Ellis:1987xu,CDF:2001kly,sonnenschein2005algebraic,betchart2014analytic,NW_D0_1999}.
A major limitation of these approaches is that experimental smearing of jet kinematics often produces equations with unphysical, complex-valued solutions.  Further, multiple real-valued solutions are inherently degenerate, requiring a specific choice of solution to be made.
Such techniques are by definition topology-specific, applicable only in systems with sufficient kinematic constraint, and fail to scale to processes with additional final-state neutrinos.
This motivates the development of alternative methods that circumvent these shortcomings.

Generative ML techniques are increasingly used in particle physics for the prediction of continuous kinematic quantities~\cite{ahmad2024comprehensive,hashemi2024deep,kansal2023evaluating}, and have recently been applied to neutrino prediction.
The $\nu$-flows model~\cite{nuflows2023} applied a conditional normalizing flow architecture~\cite{normflows2016,normflows2021} in the $t\bar{t}$ 1-lepton channel, where the normalizing flow is conditioned by a transformer encoder that takes measured event information as input.
The \nflows model~\cite{nu2flows2024} extended this method to two neutrino solutions in the $t\bar{t}$ 2-lepton channel, showing that generative ML techniques can compete with the traditional reconstruction techniques while providing solutions for all events.
More recently, diffusion architectures tackle neutrino reconstruction in $\tau \bar{\tau}$ production, where they are shown to improve the resolution of the di-tau invariant mass spectrum compared to existing non-machine learning techniques~\cite{zhang2026entanglement}.

\subsection{Towards full event reconstruction}
Algorithms for event reconstruction have generally focused on only one of the above reconstruction tasks, or have been designed with specific collider processes in mind.
Developments to the SPANet architecture introduced a regression head targeting the $t\bar{t}$ 1-lepton topology~\cite{spanet2024}. 
The model was trained to predict a single neutrino's longitudinal momentum and the invariant mass of the $t\bar{t}$ system.
This constituted a first step towards combined event reconstruction tasks, but by default only applicable to this specific process.
By fine-tuning a network pre-trained with both supervised and self-supervised tasks on a diverse set of collider processes, the EveNet foundation model authors studied combined assignment and neutrino regression downstream tasks in dileptonic $t\bar{t}$ production~\cite{hsu2026evenet}.
Alternative machine learning approaches have considered directly regressing the parent particle kinematics in $t\bar{t}$ production without decomposing the problem into individual assignment and neutrino regression components~\cite{qiu2023holistic, cms2025enhanced}.

It is evident that the feasibility and physics reach of machine learning models capable of reconstructing short-lived particles in generic collider processes has not been fully explored.
We present VyPER, the next evolution of the HyPER model, which builds upon the concept of event reconstruction using hypergraph representation learning, and is designed to be applicable to the reconstruction of arbitrary physics processes.

\section{\label{sec:methods}Methods}
VyPER is a graph-hypergraph network designed to solve both assignment and neutrino prediction tasks simultaneously.
Collider events are represented as heterogeneous graphs with a specific physics-inspired topology:
``invisible'' neutrinos are included in the graph and connected only to particles with which they share a known common parent, as outlined in Section~\ref{subsec:graph}.
The architectural backbone of VyPER is based on the message-passing technique, discussed in Section~\ref{subsec:mpnn}.
During this step, specially designed message-passing constructs learned latent representations of each neutrino.
Assignment is performed through edge and hyperedge classification, as summarized in Section~\ref{subsec:assign}.
In Section~\ref{subsec:diffusion}, we introduce the methodology for predicting neutrino kinematics: we sample random noise and map these points to physical neutrino solutions by first learning the conditional mapping between that noise and the true neutrino kinematics using the diffusion paradigm. 
The conditioning of this diffusion head relies on the neutrino latent representations constructed during the message-passing phase.
We integrate these distinct tasks into a unified loss function, enabling simultaneous optimization of all learning objectives as discussed in Section~\ref{subsec:loss}.

\begin{figure}[t]
    \includegraphics[width=0.45\textwidth]{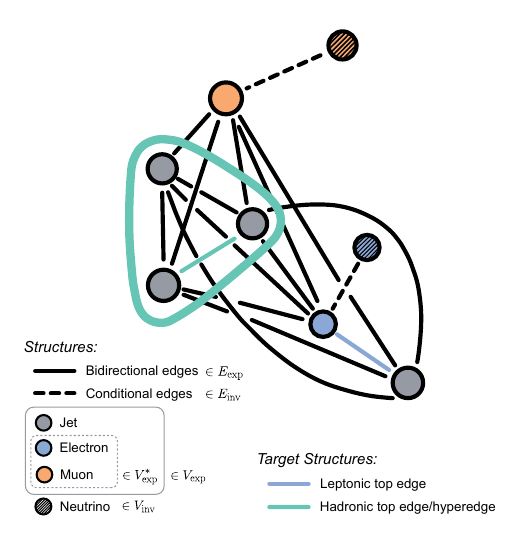}  
    \caption{\label{fig:graph_representation} VyPER graph representation of a $t\bar{t}W$ event, where the measured final state consists of four jets and two leptons of identical electric charge.
    These final states are represented as nodes and connected with edges to form a complete graph.
    The neutrinos are known to arise from leptonic $W$ boson decays and so are connected to their leptonic partner via ``conditional edges''.
    The VyPER method ultimately identifies the edges and hyperedges whose nodes correspond to the $W$ boson and top quark decay products.
    The learned neutrino representations feed into a diffusion head that predicts their kinematics.}
\end{figure}

\begin{figure*}[t]
    \includegraphics[width=1.0\textwidth]{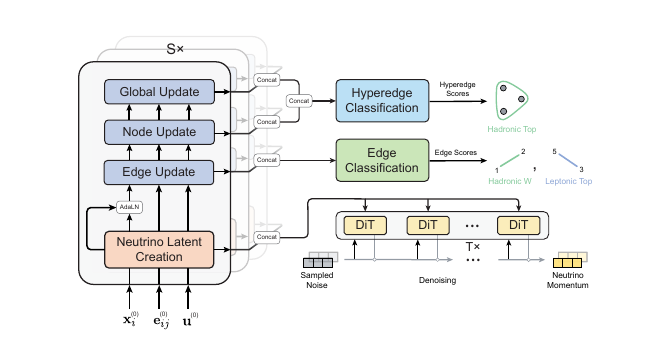}
    \caption{\label{fig:network}The VyPER network architecture. 
    The backbone is a multi-layer message-passing framework that explores and updates the graph latent sapce by exchanging information between neighboring graph objects.
    Each message-passing layer leverages relational information encoded in the graph structure to construct neutrino latent representations and update edge, node and global feature vectors.
    The updated node, edge and global states from each layer are extracted and aggregated to perform edge classification, hyperedge classification, identifying the origins of each final-state object, while the constructed neutrino representations are used to condition the reverse diffusion process, which ultimately predicts neutrino momentum.}
\end{figure*}

\subsection{\label{subsec:graph}Graph representation of collider events}
Graphs are mathematical structures composed of points and connections.
VyPER represents any collider event as a directed graph: $G=(V,E)$, where $V$ is a set of \textit{nodes} representing the final-state objects in the event, and $E$ is a set of \textit{edges} representing the directional connections between them.
Nodes are partitioned into two subsets: experimentally detectable final states $V_\mathrm{exp}$ and invisible final states $V_\mathrm{inv}$, such that $V=V_\mathrm{exp}\cup V_\mathrm{inv}$.
Edges between detectable final states are defined as $E_\mathrm{exp}=\{(i,j) \:|\: i,j\in V_\mathrm{exp}\:\mathrm{and}\: i\neq j\} \subseteq E$.
Visible final states form the subgraph $G_\mathrm{exp} (V_\mathrm{exp}, E_\mathrm{exp})$.

We define $V_\mathrm{exp}^*\subset V_\mathrm{exp}$ as the subset of visible final states that form a one-to-one correspondence with nodes in $V_\mathrm{inv}$, where each pair originates from a common parent particle.
For example, a neutrino and a charged lepton originating from the decay of a $W$ boson form such a pair.
This correspondence is defined by a bijection $f: V_\mathrm{exp}^*\xrightarrow{\textsf{\tiny bijection}} V_\mathrm{inv}$, such that their connecting edge set is defined as
\begin{equation}
    E_\mathrm{inv} = \{ (i,f(i))\:|\: i\in V_\mathrm{exp}^*\}, \label{eq:E_inv}
\end{equation}
where $f(i)\in V_\mathrm{inv}$ as the result of the bijection $f$.
Elements of $E_\mathrm{inv}$ are termed \textit{conditional edges} due to their specific role during message-passing.
In the absence of neutrinos in the final state, where $V_\mathrm{inv}=\varnothing$ and $E_\mathrm{inv}=\varnothing$, the VyPER event representation is identical to that of HyPER~\cite{hyper2024}: $G \equiv G_\mathrm{exp}$.

Following the procedure outlined in~\cite{hyper2024}, the graph $G$ is extended to a hypergraph $H(V,X$) in cases where higher-order structures are necessary: hyperedges $X$ represent resonant particles decaying into three or more final-state objects, enabling the capture of higher-order correlations among them.
For example, the hadronic decays of a top quark ($t\rightarrow bq\bar{q}'$) can be represented as an $\mathcal{O}(3)$ hyperedge.
In this case, the set of all possible hyperedges is defined as:
\begin{equation}
    X = \{\{i,j,k\} \;\vert\; i,j,k\in V_\mathrm{exp}, \; i \neq j,\; j \neq k,\; i \neq k\}.
    \label{eq:hyperedges}
\end{equation}

\subsection{\label{subsec:mpnn}Message-passing framework}
Building on advances in geometric deep learning, message-passing neural networks (MPNNs) have established themselves as one of the most effective approaches for learning graph representations~\cite{gilmer2017,battaglia2018,zhou2021,sanchezgonzalez2018}.
VyPER utilizes the message-passing framework introduced in HyPER~\cite{hyper2024}, which iteratively aggregates and propagates information through the underlying graph structure to capture the kinematic relations between final states.
In HyPER, a standard MPNN layer consisted of three sequential operations that updated the edge, node, and global attributes, respectively~\cite{hyper2024,battaglia2018}.
Extending this approach to accommodate neutrinos, VyPER leverages its graph representation (Section~\ref{subsec:graph}) to introduce an additional message-passing operation that constructs a latent representation $\tilde{\mathbf{x}}_k$ for each neutrino $k \in V_\mathrm{inv}$.
These latents are critical to the reverse diffusion process, as they condition the reconstruction of neutrino momenta (Section~\ref{subsec:diffusion}).

During each message-passing layer $s$, VyPER computes a latent representation $\tilde{\mathbf{x}}_k$ for each neutrino $k \in V_\mathrm{inv}$, by first aggregating information at its companion node, and then updating its representation based on aggregating information from the other neutrino latents:
\begin{eqnarray}
    \tilde{\mathbf{x}}^{(s)}_{i=f^{-1}(k)} &=& h^{(s)}_{1,\theta}
    \left( \mathbf{x}^{(s)}_i, \mathbf{a}^{(s)}_i, \mathbf{u}^{(s)} \right) ,
    \label{eq:tilde_x_i}
    \\
    \tilde{\mathbf{x}}^{(s)}_{i=f^{-1}(k)} &\leftarrow& h^{(s)}_{2,\theta}
    \left(
        \tilde{\mathbf{x}}^{(s)}_i ,
        \mathbf{g}^{(s)}
    \right) \;\; \mathrm{where} \; i \in V^*_\mathrm{exp} \;,
    \label{eq:tilde_x_i2}
\end{eqnarray}
where $\mathbf{x}^{(s)}_i$ contains the node features of the $i$-th node, and $\mathbf{u}^{(s)}$ is a vector of global features. 
The input event features used are defined in Appendix~\ref{apx:input}. 
Learnable functions $h^{(s)}_{\cdot,\theta}$ are implemented as Multilayer Perceptrons (MLPs), and terms in parenthesis are concatenated.
The variable $\mathbf{a}_i$ is calculated by summarizing attributes of local graph structures, including each adjacent node, $j\in V_\mathrm{exp}$, and directed edge $(i,j) \in E_\mathrm{exp}$ (denoted $j \to i$):
\begin{equation}
    \mathbf{a}^{(s)}_i =
    \bigoplus_{\forall j: \:j\rightarrow i} h^{(s)}_{3,\theta} \left(\mathbf{x}^{(s)}_i, \mathbf{x}^{(s)}_j, \mathbf{e}^{(s)}_{j\rightarrow i}\right) ,
    \label{eq:a_i}
\end{equation}
where operator $\oplus$ denotes summation,
and $\mathbf{e}^{(s)}_{j\rightarrow i}$ is an edge embedding whose inputs at $s=0$ are also given in Appendix~\ref{apx:input}.
The term $\mathbf{g}^{(s)}$ is derived from the sum of all initialized neutrino latents, so as to capture the relations between them:
\begin{equation}
    \mathbf{g}^{(s)} = h^{(s)}_{4,\theta} \left[ \bigoplus_{\forall i} (\tilde{\mathbf{x}}^{(s)}_i, \mathbf{u^{(s)}}) \right].
    \label{eq:g}
\end{equation}

To enable robust integration between the generative neutrino prediction task and the graph-based tasks of edge and hyperedge classification, the created neutrino latent vectors are propagated back to their companion nodes as messages:
\begin{eqnarray}
    \mathbf{x}^{(s)}_i \leftarrow &&\left[1 + h^{(s)}_{5,\theta}(\mathbf{x}^{(s)}_i) \right] \odot \mathbf{x}^{(s)}_i \nonumber \\
    + && \left[1+h^{(s)}_{6,\theta}(\mathbf{x}^{(s)}_i)\right] \odot \tilde{\mathbf{x}}^{(s)}_i + h^{(s)}_{7,\theta}(\mathbf{x}^{(s)}_i) \:,
    \label{eq:message_backward_pass}
\end{eqnarray}
where $\odot$ denotes element-wise product.
This technique is inspired by Adaptive Layer Normalization (AdaLN) modulation~\cite{modulation2017}, and we refer to this mechanism as \textit{conditional message-passing}.
The learned neutrino latents condition the diffusion head described in Section~\ref{subsec:diffusion}. 

After creating the neutrino latents and propagating them to the rest of the graph, the network proceeds with the remaining message-passing steps, updating the edge, node, and global features using the procedure outlined in~\cite{hyper2024}.
These message-passing operations are only applied on the sub-graph $G_\mathrm{exp}$.
The entire procedure --- conditional message-passing for neutrino latent construction and traditional message-passing to contextualize the visible graph elements --- is repeated $S$ times, where $S$ is a tunable hyperparameter.

\subsection{\label{subsec:assign} Edge and hyperedge classification}
VyPER performs edge and hyperedge classification to identify the specific set of nodes that constitute the decay products of a given parent particle.
Edge classification uses learned edge representations to pinpoint specific final-state pairs stemming from two-body decays.
Latent edge representations $\mathbf{e}_{ij}^{(s)}$ from each message-passing layer $s$ are concatenated across all $S$ message-passing steps and passed through an MLP. A softmax is then applied to give a predicted soft probability $\mathbf{e}_{ij}'\in [0,1]^{C_e}$ across $C_e$ target classes, for each edge $(i,j)\in E_\mathrm{exp}$:
\begin{equation}
    \mathbf{e}_{ij}' = \mathrm{Softmax}\left[\mathrm{MLP}\left(\concat_{s=1}^S \mathbf{e}^{(s)}_{ij}\right)\right],
\end{equation}
where $\concat$ represents concatenation.
Each class corresponds to a specific intermediate particle, and the model is trained to predict the correct class for each edge through cross-entropy loss.
The highest scoring edge within that class is selected as its reconstruction candidate. 

Classifying edges is applicable only to two-body decays. 
Hyperedge classification generalizes the approach to reconstruct parent particles with three or more decay products.
To effectively capture the multipartite correlations between final states, well-formed hyperedge latent representations are essential.
Given the set of possible hyperedges defined in Eq.~\ref{eq:hyperedges}, the embedding $\mathbf{v}_m$ for hyperedge $m \in [1,\cdots , \vert X \vert]$ is constructed by aggregating the node and global embeddings accumulated across all message-passing iterations:
\begin{equation}
    \mathbf{v}_m = \bigoplus_{l \in \varepsilon_m} \left( \mathrm{MLP}(\concat_{s=1}^S\mathbf{x}^{(s)}_l, \concat_{s=1}^S\mathbf{u}^{(s)}) \right),
    \;\;
    \forall \varepsilon_m \in X,
\end{equation}
where $\varepsilon_m$ represents the $m$-th hyperedge in $X$  and $\bigoplus_{l \in \varepsilon_m}$ denotes summation across all nodes in the hyperedge $m$.
We then compute a soft probability for each hyperedge over all classes:
\begin{eqnarray}
    \mathbf{v}_m' = \mathrm{Softmax}
    \left[
        \mathrm{MLP}(\mathbf{A}_m\odot\mathrm{ReLU(\mathbf{v}_m\mathbf{W}^T_\theta),\mathbf{v}_m})
    \right]
\end{eqnarray}
where $\mathbf{v}_m\in [0,1]^{C_X}$ is the predicted soft probability vector for hyperedge $\varepsilon_m\in X$ over $C_X$ classes, and
$\mathbf{W}^{\rm T}_\theta$ is a matrix of trainable weights.
The soft-attention coefficient vector $\mathbf{A}_m$ is computed via a per-feature softmax evaluated across all $\vert X \vert$ hyperedges:
\begin{equation}
    \mathbf{A}_m = 
    \exp{(\mathbf{v}_m)} \oslash
    \left(\bigoplus_{n=1}^{\vert X \vert} \exp{(\mathbf{v}_n)}\right),
\end{equation}
where $\oslash$ denotes element-wise division.
Each element in $\mathbf{A}_m$ lies in $[0,1]$ and represents a feature-wise importance score computed relative to all $\vert X \vert$ hyperedges.
Across all distinct classes, the hyperedge with the highest score within each class is selected as a candidate, analogous to the process of edge selection.

\subsection{\label{subsec:diffusion}Neutrino prediction through diffusion}
Generative ML architectures offer an attractive approach to predicting neutrino kinematics, as unlike models trained to regress a single kinematic solution, they can approximate the full conditional posterior distribution of neutrino kinematics for a given measured final state.
Framing neutrino prediction as generative modeling allows the network to learn a probability density over the neutrino kinematics conditioned on the measured event information, which can then be sampled to produce candidate neutrinos.

Diffusion-based approaches~\cite{ddpm2015,ddpm2020,ddim2022} are known to excel at modeling complex, multi-modal distributions while bypassing the architectural restrictions of other generative models to ensure stable training and high-fidelity predictions~\cite{dhariwal2021diffusion,croitoru2023diffusion}.
Diffusion models are defined by two complementary processes: the \textit{forward process} progressively degrades the truth-level target (the true neutrino kinematics) towards random noise, and the \textit{reverse process} employs a neural network that learns how to reverse the forward corruption.
At inference, this learned reverse process is then applied to sampled noise to generate candidate neutrino predictions.
VyPER adopts the Denoising Diffusion Implicit Model (DDIM) approach~\cite{ddim2022}, an implementation of diffusion well-known for its efficiency during inference. 

Let \(q(\mathbf{z}_0)\) denote the underlying true distribution of standardized neutrino kinematics.
Given a neutrino $\mathbf{z}_0 \sim q(\mathbf{z}_0)$, the DDIM approach defines a non-Markovian forward process that corrupts $\mathbf{z}_0$ to noise over $T$ steps. 
Instead of simulating this corruption sequentially during training, the noised neutrino latent $\mathbf{z}_t$ at arbitrary snapshot timestep $t$ can be evaluated directly in closed form using the relation
\begin{equation}
    \mathbf{z}_t = \sqrt{\bar{\alpha}_t} \mathbf{z}_0 + \sqrt{1-\bar{\alpha}_t} \: \epsilon_t ,
    \label{eq:forward_process_reparam}
\end{equation}
where $\epsilon_t$ is a sampled noise vector.
The hyperparameter $\bar{\alpha}_t = \prod_{n=1}^t \alpha_n$ controls the noise schedule; it is bounded in $(0,1]$ and monotonically decreases as $t$ increases.
In this work, the trajectory of $\bar{\alpha}_t$ is parameterized following a cosine schedule~\cite{cosine2021}.

The reverse process inverts the forward corruption defined in Eq.~\ref{eq:forward_process_reparam} by establishing a deterministic mapping from \(\mathbf{z}_{t}\) to \(\mathbf{z}_{t-1}\): 
\begin{eqnarray}
     \mathbf{z}_{t-1} &=& \sqrt{\bar{\alpha}_{t-1}} \left(\frac{\mathbf{z}_t - \sqrt{1-\bar{\alpha}_t} \epsilon_\theta(\mathbf{z}_t)}{\sqrt{\bar{\alpha}_t}}\right) \nonumber \\ 
     &&\hspace{60pt}  + \sqrt{1-\bar{\alpha}_{t-1}} \; \epsilon_\theta(\mathbf{z}_t).
     \label{eq:diffusion:ddim} 
\end{eqnarray}
where $\epsilon_\theta(\mathbf{z}_t)$ is an estimation of the noise added in the forward process.
We employ the diffusion transformer architecture~\cite{dit2023}, now standard in the generative ML literature, to predict the noise added for a given the neutrino latent at time $t$:
\begin{equation}
    \epsilon_\theta(\mathbf{z}_t) := \mathrm{DiT} (\mathbf{z_t}, \tilde{\mathbf{x}}') \:,
    \label{eq:dit}
\end{equation}
where $\tilde{\mathbf{x}}^{\prime }$ is a \textit{context  vector} encoding the measured event information.
Each neutrino has a unique context vector constructed from its latent representations built during the message-passing process (Eq.~\ref{eq:tilde_x_i} and Eq.~\ref{eq:tilde_x_i2}):
\begin{equation} 
    \tilde{\mathbf{x}}'_k = \mathrm{MLP}\left[\left(\concat_{s=1}^S \tilde{\mathbf{x}}_k^{(s)}\right), \mathbf{T}_\theta(t)\right], 
\end{equation}
where $\tilde{\mathbf{x}}_{k}^{(s)}$ represents its latent features at each message-passing step \(s\), and \(\mathbf{T}_\theta(t)\) is a timestep embedding of matching dimensionality. 
The DiT architecture extends the standard transformer with AdaLN modulation~\cite{modulation2017}, using the context vector to directly modulate the network's internal representations at every layer rather than simply concatenating the context as an additional input.
The network is trained by minimizing the mean-squared error between the predicted and ground-truth noise:
\begin{equation} 
    \mathcal{L}_\mathrm{diffusion} = \left\Vert{} \epsilon_\theta(\mathbf{z}_t) - \epsilon_t \right\Vert{}^2.  
    \label{eq:diffusion_loss} 
\end{equation}
 Upon convergence, the learned distribution \(p_\theta(\mathbf{z}_0)\) should approximate the true neutrino momentum distribution \(q(\mathbf{z}_0)\).
 At inference, the trained VyPER model starts from pure noise and uses the learned DDIM trajectories to denoise a sampled point into a realistic neutrino candidate, given the context defined by the input event features.

\subsection{\label{subsec:loss} Learning objectives}
All learning tasks are unified into a single loss function, given for each individual graph by
\begin{equation}
    \mathcal{L} = \eta \bar{\mathcal{L}}_\mathrm{dffusion} + (1-\eta) [\alpha \bar{\mathcal{L}}_\mathrm{X} 
    + (1-\alpha) \bar{\mathcal{L}}_\mathrm{E} ],
\end{equation}
where $\mathcal{L}_\mathrm{diffusion}$ is the diffusion loss formulated in Eq.~\ref{eq:diffusion_loss}; $\mathcal{L}_\mathrm{X}$ and $\mathcal{L}_\mathrm{E}$ correspond to the hyperedge and edge classification losses, respectively, computed using cross-entropy.
The bar notation, $\bar{\mathcal{L}}$, denotes averaging over the corresponding elements (neutrinos, edges or hyperedges) in each graph.
The two hyperparameters $\eta,\alpha\in[0,1]$ are loss scales for the diffusion and hyperedge components, respectively.

Training is performed with the Adam optimizer~\cite{kingma2017adammethodstochasticoptimization}.
A list of hyperparameter configurations used in the following experiments is detailed in the Appendix~\ref{apx:hparam}.

\section{\label{sec:simulation}Experimental Setup and Evaluation}

\subsection{Physics processes}
We investigate the fundamental feasibility of event reconstruction across a series of Standard Model physics processes. 
Our goal is to demonstrate that highly accurate event reconstruction is possible across diverse physics processes and final states, including those in which full reconstruction has never before been applied in the literature.
We position the VyPER model as a unified framework that can accomplish this task.

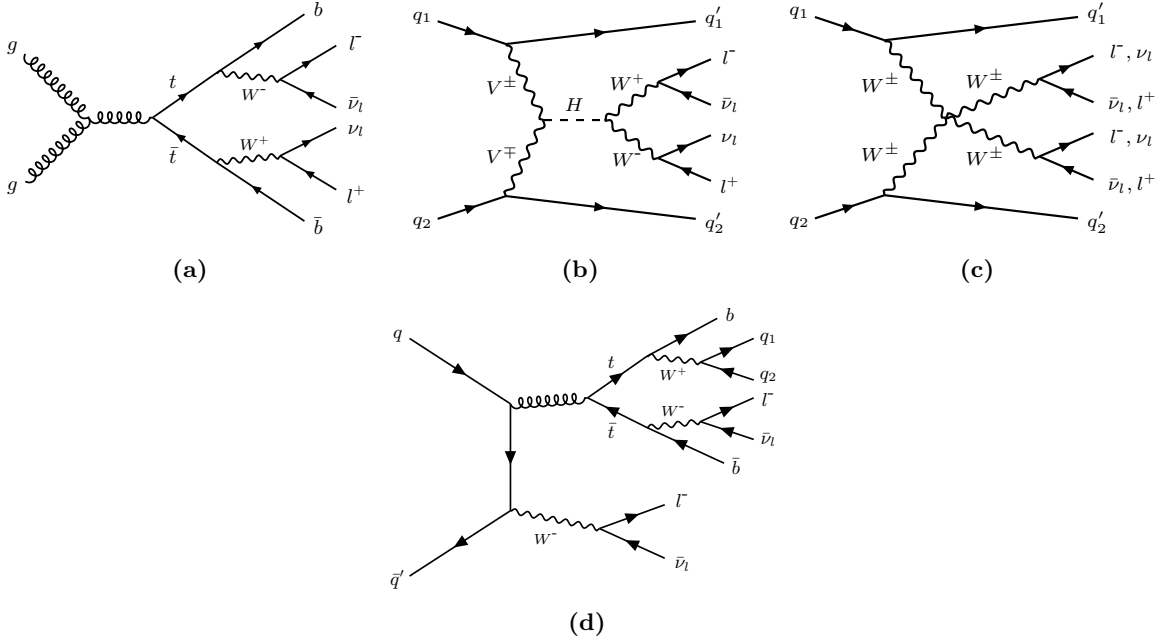
\begin{figure*}[t!]
    \centering
    \begin{tikzpicture}
        \node (diag) {\resizebox{0.28\linewidth}{!}{
\begin{tikzpicture}
\begin{feynman}
	\vertex at (0,0) (gx);
    \vertex at (-1,1) (g1);
    \vertex at (-1,-1) (g2);
    \vertex at (1,0) (gt);
    
    \vertex at (2,0.7) (t1);
    \vertex at (3.5,1.7) (b1) {};
    \vertex at (3,0.6) (wp);
    \vertex at (4.0,1.2) (q1) {};
    \vertex at (4.0,0.1) (q2) {};
    
    \vertex at (2,-0.7) (t2);
    \vertex at (3.5,-1.7) (b2) {};
    \vertex at (3,-0.6) (wm);
    \vertex at (4.0,-1.2) (q3) {};
    \vertex at (4.0,-0.1) (q4) {};
    
    \diagram[arrow size=1.0] {
        (g1) -- [gluon, line width=0.8pt] (gx);
        (g2) -- [gluon, line width=0.8pt] (gx);
        (gx) -- [gluon, line width=0.8pt] (gt);
        (gt) -- [fermion, line width=0.8pt, edge label={ $t$}] (t1);
        (t1) -- [fermion, line width=0.8pt, edge label={}] (b1);
        (t1) -- [boson, line width=0.8pt] (wp);
        (q2) -- [fermion, line width=0.8pt] (wp) -- [fermion, line width=0.8pt] (q1);
        
        (gt) -- [anti fermion, line width=0.8pt, edge label'={ $\bar{t}$}] (t2);
        (t2) -- [anti fermion, line width=0.8pt, edge label'={}] (b2);
        (t2) -- [boson, line width=0.8pt] (wm);
        (q3) -- [fermion, line width=0.8pt] (wm) -- [fermion, line width=0.8pt] (q4);
    };
\end{feynman}
 
   \draw (-1.2,+1.1) node {\small $g$};
   \draw (-1.2,-1.1) node {\small $g$};
   \draw (2.6, -0.4) node {\scriptsize $W^{\scalebox{0.7}{+}}$};
   \draw (2.6, 0.38) node {\scriptsize $W^{\scalebox{1}{-}}$};
   \draw (3.6,+1.7) node {\small $b$};
   \draw (3.6,-1.7) node {\small $\bar{b}$};
   \draw (4.2,+1.2) node {\small $l^{\scalebox{1}{-}}$};
   \draw (4.2,+0.2) node {\small $\bar{\nu}_l$};
   \draw (4.2,-1.2) node {\small $l^{\scalebox{0.7}{+}}$};
   \draw (4.2,-0.2) node {\small $\nu_l$};
\end{tikzpicture}}};
        \node[below=0mm of diag.south, font=\bfseries] {(a)};
    \end{tikzpicture}
    \begin{tikzpicture}
        \node (diag) {\resizebox{0.26\linewidth}{!}{
\begin{tikzpicture}
\begin{feynman}
	\vertex at (0,0) (vh);
	\vertex at (-0.5,1)(bosonS1);
	\vertex at (-0.5,-1)(bosonS2);
	\vertex at (0.8,0) (hw);
	\vertex at (-1.4,1.3) (initialQ1);
	\vertex at (2,1.3) (finalQ1);
	\vertex at (-1.4,-1.3) (initialQ2);
	\vertex at (2,-1.3) (finalQ2);
	\vertex at (1.5,0.5) (w1);
	\vertex at (1.5,-0.5) (w2);
	\vertex at (2.2,0.8) (w1D1);
	\vertex at (2.2,0.2) (w1D2);
	\vertex at (2.2,-0.2) (w2D1);
	\vertex at (2.2,-0.8) (w2D2);

    \diagram[arrow size=1.0] {
    	(initialQ1) -- [fermion, line width=0.8pt]  (bosonS1)  -- [fermion, line width=0.8pt] (finalQ1);
    	(initialQ2) -- [fermion, line width=0.8pt]  (bosonS2)  -- [fermion, line width=0.8pt] (finalQ2);
    	(bosonS1) -- [boson, line width=0.8pt] (vh) -- [boson, line width=0.8pt]  (bosonS2);
    	(vh) -- [scalar, line width=0.8pt, edge label={\scriptsize $H$}] (hw);
    	(hw) -- [boson, line width=0.8pt] (w1);
    	(hw) -- [boson, line width=0.8pt] (w2);
    	(w1D2) -- [fermion, line width=0.8pt]  (w1) -- [fermion, line width=0.8pt]  (w1D1); 
    	(w2D2) -- [fermion, line width=0.8pt]  (w2) -- [fermion, line width=0.8pt]  (w2D1);  
    };
\end{feynman}
 
   \draw (-1.6,+1.35) node {\scriptsize $q_1$};
   \draw (-1.6,-1.35) node {\scriptsize $q_2$};
   \draw (2.25,+1.35) node {\scriptsize $q'_1$};
   \draw (2.25,-1.35) node {\scriptsize $q'_2$};
   \draw (-0.55,-0.45) node {\scriptsize $V^\mp$};
   \draw (-0.55,+0.45) node {\scriptsize $V^\pm$};
   \draw (1.1,-0.5) node {\scriptsize $W^{\scalebox{1}{-}}$};
   \draw (1.1,+0.5) node {\scriptsize $W^{\scalebox{0.7}{+}}$};
   \draw (2.45,+0.85) node {\scriptsize $l^{\scalebox{1}{-}}$};
   \draw (2.45,+0.2) node {\scriptsize $\bar{\nu}_l$};
   \draw (2.45,-0.85) node {\scriptsize $l^{\scalebox{0.7}{+}}$};
   \draw (2.45,-0.25) node {\scriptsize $\nu_l$};
\end{tikzpicture}}};
        \node[below=0mm of diag.south, font=\bfseries] {(b)};
    \end{tikzpicture}
    \begin{tikzpicture}
        \node (diag) {\resizebox{0.29\linewidth}{!}{
\begin{tikzpicture}
\begin{feynman}
	\vertex at (0.3,0) (vs);
	\vertex at (-0.5,1)(bosonS1);
	\vertex at (-0.5,-1)(bosonS2);
	\vertex at (-1.4,1.3) (initialQ1);
	\vertex at (2,1.3) (finalQ1);
	\vertex at (-1.4,-1.3) (initialQ2);
	\vertex at (2,-1.3) (finalQ2);
	\vertex at (1.5,0.5) (w1);
	\vertex at (1.5,-0.5) (w2);
	\vertex at (2.2,0.8) (w1D1);
	\vertex at (2.2,0.2) (w1D2);
	\vertex at (2.2,-0.2) (w2D1);
	\vertex at (2.2,-0.8) (w2D2);

    \diagram[arrow size=1.0] {
    	(initialQ1) -- [fermion, line width=0.8pt]  (bosonS1)  -- [fermion, line width=0.8pt] (finalQ1);
    	(initialQ2) -- [fermion, line width=0.8pt]  (bosonS2)  -- [fermion, line width=0.8pt] (finalQ2);
    	(bosonS1) -- [boson, line width=0.8pt] (vs) -- [boson, line width=0.8pt]  (bosonS2);
    	(w1) -- [boson, line width=0.8pt]  (vs) -- [boson, line width=0.8pt] (w2);
    	(w1D2) -- [fermion, line width=0.8pt]  (w1) -- [fermion, line width=0.8pt]  (w1D1); 
    	(w2D2) -- [fermion, line width=0.8pt]  (w2) -- [fermion, line width=0.8pt]  (w2D1);  
    };
\end{feynman}
 
   \draw (-1.6,+1.35) node {\scriptsize $q_1$};
   \draw (-1.6,-1.35) node {\scriptsize $q_2$};
   \draw (2.25,+1.35) node {\scriptsize $q'_1$};
   \draw (2.25,-1.35) node {\scriptsize $q'_2$};
   \draw (-0.55,-0.45) node {\scriptsize $W^\pm$};
   \draw (-0.55,+0.45) node {\scriptsize $W^\pm$};
   \draw (0.8,-0.5) node {\scriptsize $W^\pm$};
   \draw (0.8,+0.5) node {\scriptsize $W^\pm$};
   \draw (2.7,+0.85) node {\scriptsize $l^{\scalebox{1}{-}},\nu_l$};
   \draw (2.7,+0.2) node {\scriptsize $\bar{\nu}_l,l^{\scalebox{0.7}{+}}$};
   \draw (2.7,-0.85) node {\scriptsize $\bar{\nu}_l,l^{\scalebox{0.7}{+}}$};
   \draw (2.7,-0.25) node {\scriptsize $l^{\scalebox{1}{-}},\nu_l$};
\end{tikzpicture}}};
        \node[below=0mm of diag.south, font=\bfseries] {(c)};
    \end{tikzpicture}
    \begin{tikzpicture}
        \node (diag) {\resizebox{0.3\linewidth}{!}{
\begin{tikzpicture}[
  every node/.style = {font=\small}, 
  every edge label/.style = {font=\small} ]

 \begin{feynman}
 	 \vertex at (0,+2) (hard_initial_q1) ;
     \vertex at (0,-2) (hard_initial_q2) ;
     \vertex at (1.7,0.9) (hard_med_g_i) ;
     \vertex at (3., 1.0) (hard_med_g_j) ;
     \vertex at (1.7,-0.9) (hard_med_W_i) ;
     \vertex at (3.2,-1.2) (hard_med_W_j) ;
     \vertex at (4.3,-0.8) (hard_med_W_dec_q1) ;
     \vertex at (4.3,-1.7) (hard_med_W_dec_q2) ;
     \vertex at (4., 1.7) (hard_med_t_i) ;
     \vertex at (4., 0.5) (hard_med_tbar_i) ;
     \vertex at (5.3, 2.4) (hard_dec_b){} ;
     \vertex at (5.3, -0.1) (hard_dec_bar) ;
     \vertex at (4.9, 1.6) (hard_dec_wp) ;
     \vertex at (4.9, 0.6) (hard_dec_wm) ;
     \vertex at (5.8, 2.0) (hard_dec_wp_dec_q1) ;
     \vertex at (5.8, 1.3) (hard_dec_wp_dec_q2) ;
     \vertex at (5.8, 1.0) (hard_dec_wm_dec_q1) ;
     \vertex at (5.8, 0.3) (hard_dec_wm_dec_q2) ;

    \diagram*{
        (hard_initial_q1) -- [fermion, line width=0.8pt] (hard_med_g_i) -- [fermion, line width=0.8pt] (hard_med_W_i) -- [fermion, line width=0.8pt] (hard_initial_q2),
     	(hard_med_g_i) -- [gluon, line width=0.8pt] (hard_med_g_j),
        (hard_med_W_i) -- [boson, line width=0.8pt](hard_med_W_j),
        (hard_med_g_j) -- [fermion, line width=0.8pt](hard_med_t_i) -- [fermion, line width=0.8pt](hard_dec_b),
        (hard_med_g_j) -- [anti fermion, line width=0.8pt](hard_med_tbar_i) -- [anti fermion, line width=0.8pt](hard_dec_bar),
        (hard_med_t_i) -- [boson, line width=0.8pt](hard_dec_wp),
        (hard_med_tbar_i) --[boson, line width=0.8pt](hard_dec_wm),
        (hard_dec_wp_dec_q2) -- [fermion, line width=0.8pt](hard_dec_wp) -- [fermion, line width=0.8pt](hard_dec_wp_dec_q1) ,
        (hard_dec_wm_dec_q2) -- [fermion, line width=0.8pt](hard_dec_wm) -- [fermion, line width=0.8pt](hard_dec_wm_dec_q1) ,
        (hard_med_W_dec_q2) -- [fermion, line width=0.8pt](hard_med_W_j) -- [fermion, line width=0.8pt](hard_med_W_dec_q1) ,
     };
     
    \node[font=\scriptsize] at (2.3,-1.35) {$W^{\scalebox{1.0}{-}}$};
    \node[] at (4.6,-0.7) {$l^{\scalebox{1.0}{-}}$};
    \node[] at (4.6,-1.8) {$\bar{\nu}_l$};
    \node[] at (3.4,1.6) {$t$};
    \node[] at (3.4,0.45) {$\bar{t}$};
    \node[font=\scriptsize] at (4.45,1.4) {$W^{\scalebox{0.7}{+}}$};
    \node[font=\scriptsize] at (4.45,0.8) {$W^{\scalebox{1.0}{-}}$};
    \node[] at (5.4,2.4) {$b$};
    \node[] at (6.05,2) {$q_1$};
    \node[] at (6.05,1.35) {$q_2$};
    \node[] at (6.05,1) {$l^{\scalebox{1.0}{-}}$};
    \node[] at (6.05,0.3) {$\bar{\nu}_l$};
    \node[] at (5.5,-0.15) {$\bar{b}$};
    \node[] at (-0.2,2.05) {$q$};
    \node[] at (-0.2,-2.05) {$\bar{q}'$};
\end{feynman}
\end{tikzpicture}}};
        \node[below=0mm of diag.south, font=\bfseries] {(d)};
    \end{tikzpicture}    
    \caption{Representative diagrams at leading order in both QCD and electroweak couplings for the two-lepton final states of (a)  \(t\bar{t}\), (b)  VBF \(H\rightarrow WW^*\), (c) same-sign VBS \(WW\), and (d) same-sign dileptonic \(t\bar{t}W\).}
    \label{fig:feynmans}
\end{figure*}

We consider four distinct physics processes. 
First, the production of top-antitop quark pairs decaying into the dileptonic final state, \(t\bar{t}(2\mathrm{L})\), allows us to benchmark machine learning models against established analytical reconstruction algorithms. 
Second, the production of \(W\) boson pairs in the electroweak sector — via either the decay of a Higgs boson produced through vector-boson fusion (\(H \rightarrow WW^*\)) or vector-boson scattering (VBS \(WW\)) — demonstrates how machine learning achieves full event reconstruction in under-constrained channels.
Finally, the rare and complex production of a $W$ boson in association with a top-antitop quark pair ($t\bar{t}W$) combines multiple combinatoric assignment tasks with neutrino prediction, offering a highly compelling challenge for modern event reconstruction methods.
Feynman diagrams for all four processes are shown in Figure~\ref{fig:feynmans}.
These channels are chosen because they prioritize different aspects of the reconstruction problem and span a wide range of kinematic constraints, from the \(t\bar{t}\) system with all resonances on their mass shells, to the highly under-constrained VBS \(WW\) topology.
Successfully reconstructing under-constrained topologies will demonstrate the capacity of ML models to learn implicit physics constraints and resolve under-determined systems that are algebraically intractable.

We perform all studies using simulated data, where the generator-level truth record allows us to directly evaluate reconstruction performance through computing the efficiency of jet and charged lepton assignments, and by evaluating the similarity of predicted kinematic spectra to their true counterparts.

\begin{table}[!ht]
\caption{Overview of the four processes studied, detailing the channel and basic event selections, and the number of events used for training, testing and validation of all ML-based models.``2LOS" indicates the requirement of two leptons of opposite electric charge in the final state;
``2LSS" indicates that both leptons have the same electric charge. 
}
\label{tab:processes}
\begin{ruledtabular}
\begin{tabular}{cccccc}
Process     & $N_\mathrm{jet}$ & $N_{b\text{-jet}}$ & Channel & Split & $N_\mathrm{events}$ \\[1pt]\hline\\[-7pt]
\multirow{3}{*}{$t\bar{t}$ 2L} & \multirow{3}{*}{$\geq$2} & \multirow{3}{*}{$\geq$2} & \multirow{3}{*}{2LOS} & Train & 4,692,288  \\
& & & & Val. & 260,190 \\
& & & & Test & 261,557 \\[1pt]\hline\\[-7pt]
\multirow{3}{*}{$H\mkern-5mu\to\mkern-5mu WW^*$} & \multirow{3}{*}{$\geq$2} & \multirow{3}{*}{$\geq$0} & \multirow{3}{*}{2LOS} & Train & 4,765,195  \\
& & & & Val. & 266,844 \\
& & & & Test & 266,919 \\[1pt]\hline\\[-7pt]
\multirow{3}{*}{VBS $WW$} & \multirow{3}{*}{$\geq$2} & \multirow{3}{*}{$\geq$0} & \multirow{3}{*}{2LSS} & Train & 4,145,532  \\
& & & & Val. & 107,179 \\
& & & & Test & 106,491 \\[1pt]\hline\\[-7pt]
\multirow{3}{*}{$t\bar{t}W$ 2L} & \multirow{3}{*}{$\geq$4} & \multirow{3}{*}{$\geq$0} & \multirow{3}{*}{2LSS} & Train & 4,907,261  \\
& & & & Val. & 272,740 \\
& & & & Test & 272,235 \\
\end{tabular}
\end{ruledtabular}
\end{table}

\subsection{Simulation and event selection}
All physics processes are simulated at a center-of-mass energy of $\sqrt{s}=$ 13 TeV.
The hard-scattering matrix elements are evaluated at next-to-leading-order (NLO) accuracy in quantum chromodynamics (QCD) and leading-order (LO) accuracy in the electroweak (EW) coupling using the \textsc{MadGraph5}\_aMC@NLO framework (v3.5.7-LTS)~\cite{Alwall:2014hca}.
The NLO five-flavour scheme parton distribution function (PDF) set of \textsc{NNPDF3.0nlo}~\cite{NNPDF:2014otw} is used.
Decays of the top quark, $W$ boson and Higgs boson are modeled using \textsc{MadSpin}~\cite{Artoisenet:2012st}.
Events are matched to \textsc{Pythia8} (v8.313)~\cite{Bierlich:2022pfr} to simulate parton shower, hadronization, and multi-parton interactions.
Detector response is simulated using \textsc{Delphes} (v3.5.0)~\cite{deFavereau:2013fsa}, configured to be similar to that of the ATLAS detector.

Jets are reconstructed using the anti-$k_t$ algorithm~\cite{Cacciari:2008gp} with a radius parameter of $R=0.4$, as implemented in \textsc{FastJet} (v3.4.0)~\cite{Cacciari:2011ma}.
Jets are required to have a minimum $p_T$ of 25 GeV, and a minimum of 10 GeV is required for electrons and muons.
Jets and electrons must satisfy $|\eta|\leq2.5$, and $|\eta|\leq2.7$ is required for muons.
The identification of jets originating from a B-hadron ($b$-tagging) is performed by \textsc{Delphes} with a $p_T$-dependent tagging efficiency based on \cite{ATL-PHYS-PUB-2015-022}.

All machine learning techniques tested employ supervised learning tasks that rely on truth information for model training and evaluation of reconstruction performance.
The momenta of the parton-level truth objects, such as top quarks, Higgs bosons, $W$ bosons, and their decays, are extracted directly from the \textsc{Pythia8} simulation record after QCD radiation. 
Particle assignment labels are established by matching reconstructed detector objects to generator-level partons --- the immediate products of parent parton decays --- based on the angular distance metric $\Delta R = \sqrt{(\Delta\phi)^2+(\Delta\eta)^2}$.
A jet is considered to have originated from a particular parton if the angular distance between them satisfies $\Delta R < 0.4$.
Lepton matching is performed with a $\Delta R$ threshold of 0.1.
In cases where the detector-level lepton multiplicity exceeds that at truth level, the leading detector-level leptons are selected according to the truth-level multiplicity before matching.

Table~\ref{tab:processes} lists the set of basic selections applied to each dataset to isolate the desired final state.
We focus on final states with two leptons, as these decay channels provide multiple neutrinos that present a testbed for neutrino momentum prediction.
Dileptonic final states can be characterized by the electric charge of the leptons.
Both \(t\bar{t}\) and \(H \rightarrow WW^*\) processes produce parent particles with opposite electric charge, giving a final state that has two oppositely charged leptons (2LOS).
We study VBS $WW$ production and $t\bar{t}W$ production in the channel with two leptons of identical electric charge: the two-lepton same-sign (2LSS) channel.

\subsection{Training and evaluation}
The simulated datasets are split into independent training, validation, and testing sets, with the first two used for model training and in-training validation, respectively. 
Performance evaluation is conducted on the testing datasets; for processes where traditional, non-ML methods are available, they are likewise evaluated only on this common testing set.
The number of events dedicated to training, validation and testing are quoted in Table~\ref{tab:processes}.
Separate models are trained independently for each process\footnote{Studies of transfer learning between processes is left as future work and to dedicated foundation models~\cite{hsu2026evenet}.}.
All datasets have been made publicly available \cite{mao_2026_22308461} as outlined in Appendix~\ref{sec:soft_data_avail}.

\subsection{\label{subsec:metrics}Performance metrics}
The performance of an event reconstruction algorithm is inherently tied to the specific requirements of a given physics measurement. 
In an attempt to remain agnostic to any single downstream analysis, we evaluate our algorithms using a generic suite of performance markers.

\paragraph{Assignment efficiency:}
To quantify the performance of the assignment algorithms, we define assignment efficiency as the fraction of correctly reconstructed parent particles relative to the total number of target parents.
A hadronic top quark is deemed correctly reconstructed if and only if both the $b$-jet and the $W$ boson are also correctly assigned (with the permutation of the two jets in the $W$ boson irrelevant in the HyPER assignment paradigm).

\paragraph{Neutrino residuals:}
To quantify the performance of the neutrino prediction methods, we evaluate event-by-event differences between the true neutrino kinematics and the predictions of each algorithm. 
Specifically, we present the mean \(\Delta R\) separation between the true and predicted vectors, alongside the root mean square error (RMSE) of the Cartesian momentum components. 
We also compute the residuals for each momentum component (e.g., \(p_x^{\text{Truth}} - p_x^{\text{Pred}}\)) and extract the ``effective interquantile resolution'' (Res.) between the 15.87-th and 84.13-th percentiles. 
These specific quantiles are chosen because they bound the central 68.27\% of the data, mapping directly to the standard \(\pm1\sigma\) intervals of a normal distribution to yield a \(2\sigma\) total width for the reported resolution.
All four final states contain two neutrinos; consequently, the distributions are produced on a per-neutrino basis rather than per-event.
For all evaluated metrics, smaller values indicate closer proximity to the truth and so superior reconstruction performance.

\paragraph{Parent particle distributions:}
Distributions constructed from reconstructed parent particle
kinematics. We compare to “idealized reconstruction” distributions which are detector-level quantities built using detector-level object kinematics
with true assignments and the true neutrino kinematics. The idealized reconstruction distributions
represent the limiting case of perfect event reconstruction, and showcase each method’s ability to recover the shape of a particular observable.

\paragraph{High-level observable comparison to truth:}
An event-by-event comparison between the prediction from each algorithm for a specific reconstructed observable, and a predefined truth definition of the observable as recorded in the simulation record.
This can be rendered in one-dimension as a plot of the residuals, analogously to the neutrino kinematic residuals, or in two dimensions as a binned migration matrix which demonstrates the correlation and migration of events between detector-level and truth-level.
 We again employ the effective interquantile resolution (Res.) between the 15.87-th and 84.13-th percentiles as a summary metric in the former case.
We use the \textit{Trace} to summarize the latter case: the proportion of normalized event counts falling along the leading diagonal of the matrix.
This represents simply the proportion of events reconstructed in the correct bin, with a higher value indicating a more diagonal migration matrix.
Per convention, migration matrices are normalized row-by-row, and as such the trace is normalized to unity by dividing by the number of rows.

In the event-by-event comparison of high-level observables, the definition of ``truth'' varies. In the $t\bar{t}$ and $t\bar{t}W$ studies, we compare to a \textit{partonic} truth definition: the top-quark or $W$ boson kinematics as recorded in simulation immediately prior to their decay, preceding the application of parton showering, hadronization, and subsequent detector response modeling.
We choose this because measurements in the top-quark sector frequently unfold to an inclusive partonic definition for comparison to fixed-order perturbative QCD predictions.
In the electroweak studies, we use the idealized reconstruction definition as a proxy for a stable, ``particle-level'' fiducial phase-space.
This is valid as the observables in question are built exclusively from charged leptons and neutrinos, and as such do not suffer from large smearing effects arising from the detector response to jets.

Bold typeface in tabulated results denotes the best performance for each metric, with columns corresponding to individual observables.
Neutrino residual distributions and migration matrices are plotted exclusively for the \(t\bar{t}\)(2L) process, along with their corresponding tabulated summary metrics. 
For the remaining processes, only the tabulated summaries are provided.
Uncertainties are computed for RMSEs, traces and effective resolutions by bootstrapping 100 times and taking the standard deviation of the results. 
The computed uncertainties are found to be uniformly smaller than the differences between reconstruction techniques, thus quoted variations in performance between methods are statistically significant. 
To maintain readability, individual uncertainties are omitted from tables, and the maximum observed uncertainty is quoted in each table caption.

\section{\label{sec:tt2L} Event reconstruction in dileptonic $t\bar{t}$}
With its characteristic short lifetime and chiral decay mediated by the $W$ boson, the top quark exhibits a rich phenomenology that may be studied with high precision in $t \bar{t}$ production at the LHC.
Measurements of differential cross-sections, unfolded into the full partonic phase-space, provide invaluable comparisons for fixed-order calculations of $t\bar{t}$ production and for developments in event generation~\cite{sirunyan2017measurement,catani2019top}.
The spin polarizations of the individual quarks and the correlations between them are probed through precise measurement of angular observables in $t\bar{t}$ production~\cite{bernreuther2015set,sirunyan2019measurement}, and offer insight into the presence of quantum entanglement and bound-state effects near the production threshold~\cite{atlas2024observation,atlas2026observation_toponium}.
Kinematic quantities pertaining to the top quarks and the overall $t\bar{t}$ system set constraints on various supersymmetric (SUSY) scenarios~\cite{sirunyan2018search}, extended Higgs models~\cite{atlas2025search} and Standard Model effective field theory (SMEFT) operators~\cite{saavedra2018interpreting}.
All scenarios require accurate reconstruction of the top quarks' kinematics, demonstrating why top quark reconstruction has been an active area of research since the Tevatron~\cite{NW_D0_1999}.

\begin{figure*}[t!]
    \centering
    \includegraphics[width=0.98\textwidth]{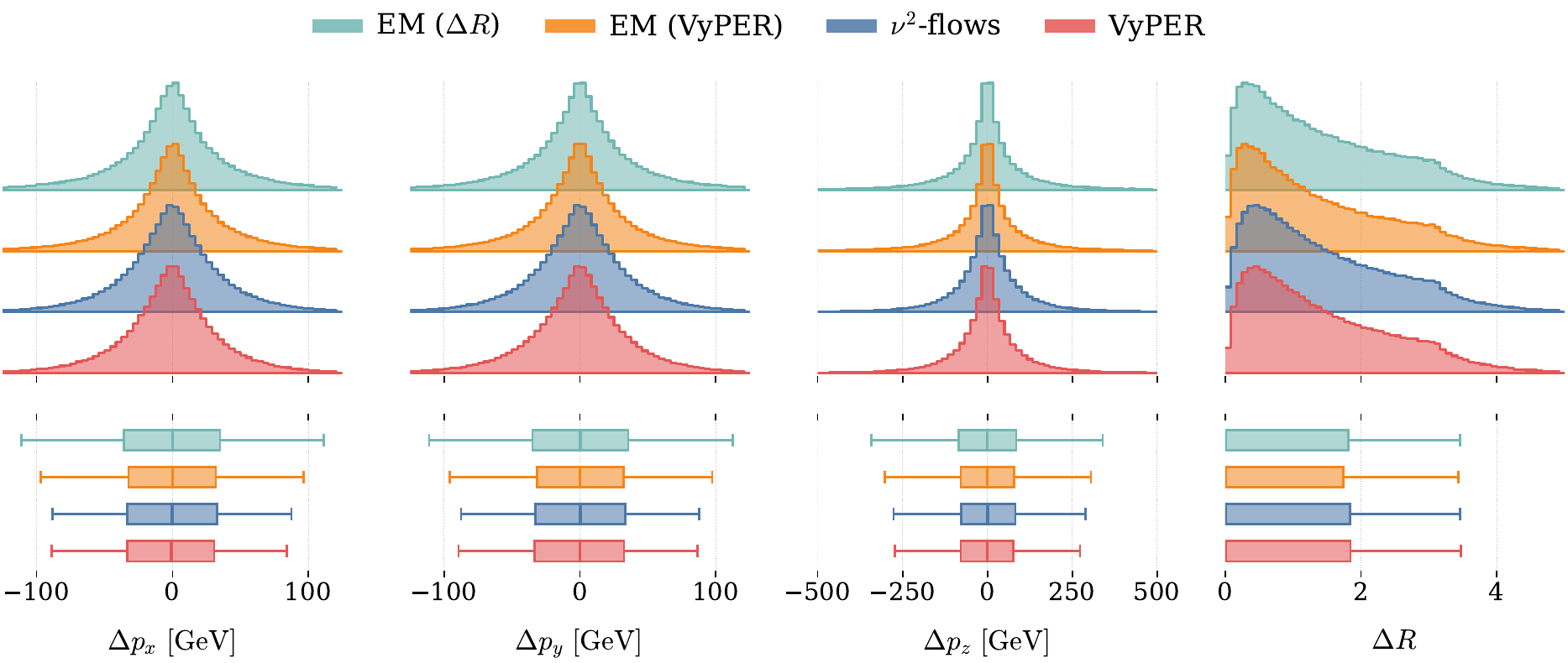}
    \caption{\label{fig:ttbar_neutrino_res}
    Binned distributions between the truth neutrino and predicted neutrino Cartesian momentum ($p_x$, $p_y$, $p_z$), as well as their angular separation $\Delta R$, are shown in the upper plot.
The lower box plots define the spread around the median (central line) for \(p_{x}\), \(p_{y}\), and \(p_{z}\) using the \(\pm 1\sigma\) and \(\pm 2\sigma\) quantiles (corresponding to the 15.87th–84.13th and 2.28th–97.72th percentiles, respectively). 
The strictly positive \(\Delta R\) distribution is summarized using a one-sided box plot, with intervals corresponding to \(68.3\%\) and \(95.4\%\) containment to represent \(1\sigma\) and \(2\sigma\) coverage.}
\end{figure*}

\subsection{\label{subsec:tt2L_ellipse} Analytical ellipse method}
By imposing constraints on the known masses of the top quark and \(W\) boson, the system of kinematic equations relating the \(t\bar{t}\) system to its final-state decay products can be closed and solved analytically. 
Several non-ML-based techniques have been developed specifically for reconstructing neutrinos in dileptonic $t\bar{t}$~\cite{sonnenschein2005algebraic,betchart2014analytic,NW_D0_1999}.
We focus on the Ellipse Method (EM)~\cite{ellipse2014}, which has been utilized in several recent ATLAS and CMS $t\bar{t}$ measurements~\cite{ellipse_CMS_2019,atlas2024observation,atlas2026observation_toponium,khachatryan2017measurement},
and frames the system of equations geometrically. 
For each top or antitop quark decay chain, the \(W\)-boson mass constraint confines the unobserved neutrino momentum to a two-dimensional ellipse within the transverse momentum plane. The intersections of the two ellipses --- representing the separate top and antitop decay branches --- yield the physically permissible momentum solutions for the neutrino system.

Our implementation\footnote{\url{https://github.com/els285/VyPERpaper}} of the EM extends the framework provided in~\cite{ellipse2014} by incorporating a stochastic sampling of the top-quark and \(W\)-boson mass distributions. Performing multiple mass samplings per event allows the algorithm to make multiple attempts to find analytical solutions, thereby enhancing the reconstruction efficiency. In events where multiple valid solutions are found, the configuration yielding the lowest invariant mass of the $t\bar{t}$ system ($m_{t\bar{t}}$) is selected. The top-quark mass is sampled from a Gaussian distribution $\mathcal{N}(172.50\text{ GeV}, 1.48\text{ GeV})$, while the \(W\)-boson mass is drawn from $\mathcal{N}(80.38\text{ GeV}, 2.085\text{ GeV})$.

Alternative analytical solution procedures utilized in past measurements include the Sonnenschein \cite{sonnenschein2005algebraic} and NW methods \cite{NW_D0_1999}, which are described in Appendix~\ref{sec:ttbar_alt}.
The Sonnenschein method has been reported to perform worse than the EM \cite{Simpson:2024hbr} and was excluded from all studies, as was the case in the $\nu^2$-flows publication \cite{nu2flows2024}.
A custom implementation\footnote{\url{https://github.com/tzuhanchang/TensorNW}} of the NW method was evaluated on our test dataset but dropped from final comparison plots due to its inferior reconstruction performance relative to the EM.

\subsection{Assignment performance}
The assignment task in dileptonic $t\bar{t}$ is simple: associate a unique $b$-jet to each charged lepton to define the visible decay products of both parent quarks.
We study how SPANet and VyPER perform in this task.
The specific procedure for VyPER is given in Appendix~\ref{apx:assign-strategy}. 
Table~\ref{tab:ttbar_efficiency} shows this comparison.
The per-top assignment efficiency $\varepsilon{(t_\mathrm{lep})}$ is the ratio of correctly assigned $b$-jet--lepton pairs to the total number of top quarks, and the event assignment efficiency $\varepsilon({t \bar{t}})$ is the ratio of events in which both pairings are correct to the total number of events.
The SPANet training setup is discussed in Appendix~\ref{sec:appendix_nu2flows_spanet}.
The assignment performance is comparable between both networks.

\begin{table}[!t]
    \caption{The assignment efficiencies, $\varepsilon=N_\mathrm{correct}/N_\mathrm{total}$, for the individual leptonic top quark, $t_{\rm lep}$, and the correct pair, $t\bar{t}$, for VyPER and SPANet.
    The absolute uncertainty on $\varepsilon_{t_{\rm lep}}^{\quad(\%)}$ is 0.05\% and 0.07\% for $\varepsilon_{t\bar{t}}^{\quad(\%)}$, for both methods.}
    \label{tab:ttbar_efficiency}
    \begin{ruledtabular}
    \begin{tabular}{lcc}
    Model & $\varepsilon_{t_{\rm lep}}^{\quad(\%)}$  & $\varepsilon_{t\bar{t}}^{\quad(\%)}$ \\[3pt]\hline\\[-7pt]
    VyPER  & \textbf{85.0} & \textbf{83.8}  \\[3pt]
    SPANet & 84.7          & 83.4       \\
    \end{tabular}
    \end{ruledtabular}
\end{table}


\subsection{Neutrino reconstruction}
The VyPER, $\nu^2$-flows and EM approaches are used to predict the neutrino kinematics.
While the machine learning (ML) methods utilize all measured event kinematics as inputs, the EM relies on predetermined \(b\)-jet and lepton assignments. To evaluate its performance, the EM is tested in two configurations. The first utilizes assignments directly provided by the VyPER assignment head, and is referred to as ``EM(VyPER)''. The second employs a legacy geometric approach mimicking historic ATLAS and CMS implementations without ML inputs; this configuration selects the two highest-\(p_{T}\) \(b\)-tagged jets in the event and uses \(\Delta R\) matching to assign the \(b\)-jet closer to the positively charged lepton to the top quark decay.
It is referred to as ``EM($\Delta R$)''.
The $\nu^2$-flows implementation and hyperparameters are discussed in Appendix~\ref{sec:appendix_nu2flows_spanet}.

A major limitation of the analytical EM method is its inability to yield physically valid solutions for a non-negligible fraction of events. 
Table~\ref{tab:tt2L_EM_eff} presents the reconstruction efficiency of the EM solver when the $b$-jet--lepton pairings are determined via legacy \(\Delta R\) geometric matching versus the machine-learning-driven VyPER assignment head. 
A significant recovery in efficiency is achieved when utilizing the VyPER assignment, demonstrating substantial performance gains realized by augmenting traditional analytical solvers with a machine-learning-based assignment.
The performance gains realized from sampling the top quark and  \(W\) boson masses 20 times are quantified in the second row of Table~\ref{tab:tt2L_EM_eff}.

\begin{table}[!t]
\centering
\caption{The fraction of events for which the Ellipse Method provides real solutions, for cases where a $\Delta R$ assignment of $b$-jet and lepton is given as input, and where this assignment is taken from VyPER.
The effect of repeated sampling of the $W$ boson and top quark mass distributions per event is also shown.
The uncertainty on the efficiency from the $\Delta R$ approach is 0.07\%, and 0.05\% for the VyPER approach.}
\label{tab:tt2L_EM_eff}
\begin{ruledtabular}
\begin{tabular}{lcc}
Mass sampling & EM($\Delta R$) [\%] & EM(VyPER) [\%] \\[1pt] \hline\\[-7pt]
1   & $83.9  $ & $\textbf{92.1} $ \\
20  & $87.0 $  & $\textbf{95.0} $ \\
\end{tabular}
\end{ruledtabular}
\end{table}

\begin{figure*}[!t]
    \centering
    \includegraphics[width=\textwidth]{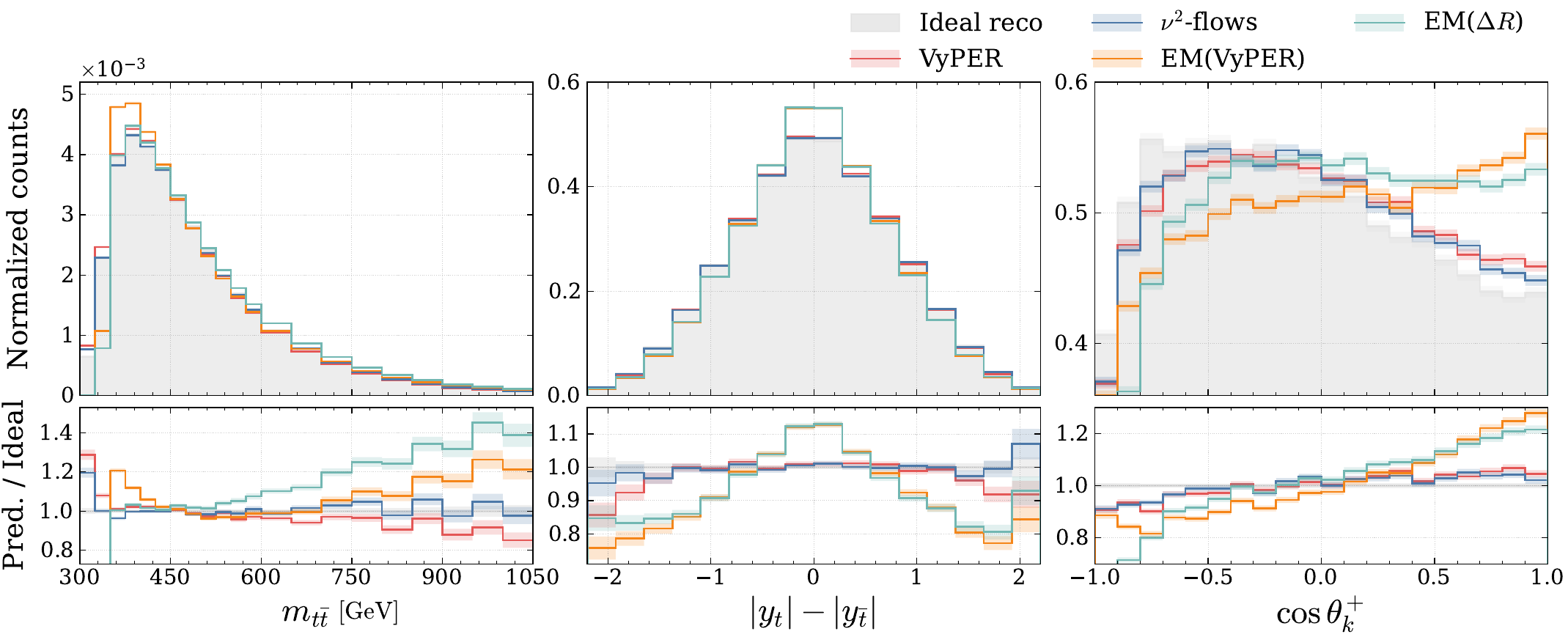}
    \caption{ \label{fig:tt2L_distributions_row}
    Distributions of observables in \(t\bar{t} (2L)\) production  reconstructed using VyPER, $\nu^2$-flows with VyPER assignment, EM with VyPER assignment, and EM with a historic $\Delta R$ assignment. 
    Each prediction is compared to the idealized reconstruction target.
    Distributions are normalized to unit integral to account for differences in the number of events, as EM(VyPER) and EM($\Delta R$) contain only a subset of events that have real solutions.
    Shaded bands indicate statistical uncertainty in each bin.}
\end{figure*}

\begin{figure*}[!t]
    \centering
    \includegraphics[width=\textwidth]{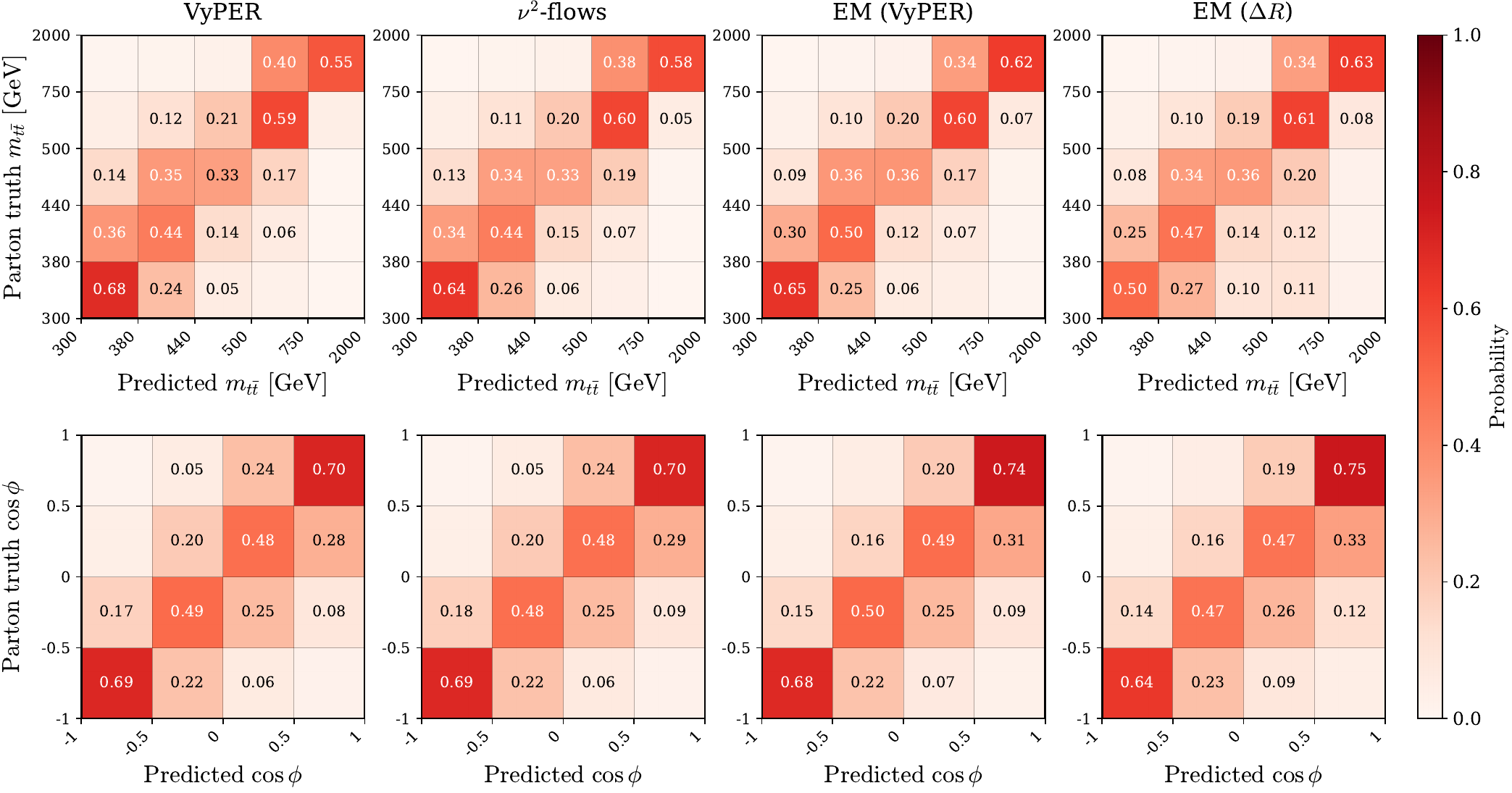}
    \caption{Migration matrices for $m_{t\bar{t}}$ and $\cos \phi$ observables for the four reconstruction approaches in the $t\bar{t} $(2L) process.
    The total count of each row is normalized to unity by convention, and the fractional per-row yield is annotated in each bin. 
    Bins with under 5\% of the row's yield are not annotated.
    The binning for the $m_{t\bar{t}}$ matrices is not equal, with finer binning concentrated around the bulk of the distribution, but presented with equal spacing for readability.}
    \label{fig:tt2L_migrations}
\end{figure*}

\begin{table}[!t]
\caption{Neutrino reconstruction accuracy in $t\bar{t}$ 2L production.
Metrics defined in Section~\ref{subsec:metrics}. 
Momentum metrics are quoted in GeV.
The relative uncertainty does not exceed 0.4\% on the Res. and does not exceed 0.75\% on the RMSE.
Uncertainties are larger on EM predictions and highest for the EM($\Delta R$) method, driven by a reduction in statistics from events failing the reconstruction.}
\label{tab:tt2L_nu_residuals}
\begin{ruledtabular}
\begin{tabular}{lccccccc}
& $\Delta R$ & \multicolumn{2}{c}{$\Delta p_x$} & \multicolumn{2}{c}{$\Delta p_y$} & \multicolumn{2}{c}{$\Delta p_z$} \\
\cmidrule(lr){2-2} \cmidrule(lr){3-4} \cmidrule(lr){5-6} \cmidrule(lr){7-8}
Model & Mean & Res. & RMSE & Res. & RMSE & Res. & RMSE\\[1pt]\hline\\[-7pt]
EM($\Delta R$) & 2.24 & 70.6 & 54.4 & 70.3 & 54.6 & 170 & 170 \\
EM(VyPER)          & 2.20 & 64.3 & 46.2 & \textbf{63.9} & 46.1 & 158 & 149 \\
$\nu^2$-flows  & \textbf{2.19} & 66.3 & 40.9 & 66.1 & \textbf{40.6} & 156 & 133 \\
VyPER          & 2.20 & \textbf{64.1} & \textbf{39.8} & 66.0 & \textbf{40.6} & \textbf{155} & \textbf{127} \\
\end{tabular}
\end{ruledtabular}
\end{table}

Figure~\ref{fig:ttbar_neutrino_res} presents the neutrino kinematic residuals --- defined as the true value less the predicted --- for \(p_{x}\), \(p_{y}\), \(p_{z}\), 
alongside the angular separation between the true and reconstructed neutrinos. 
We use the results obtained from the 20-fold mass smearing for both  EM(\(\Delta R\)) and  EM(VyPER) methods, and events that fail to yield a solution are excluded from that method's residuals. 
Conversely, VyPER and \(\nu ^{2}\)-flows use the full dataset. 
All histograms are normalized such that their integral is unity.
In the bottom panels, the box plots illustrate data distribution using standard deviation intervals around the median for the Cartesian momenta. The inner boxes capture the \(\pm 1\sigma\) range (15.87-th to 84.13-th percentiles), while the outer whiskers extend to the \(\pm 2\sigma\) range (2.28-th to 97.72-th percentiles, containing 95.5\% of the data).
The $\Delta R$ displays the median, and the  \(1\sigma\), and  \(2\sigma\) deviation from zero.
The ML techniques exhibit slightly narrower distributions for the momentum components.
The deviation from zero for the neutrino's \(p_{z}\) momentum is much larger than its transverse components because the MET provides no longitudinal constraints.
The effective resolution (Res.) and the RMSE for each distribution and technique are also tabulated in Table~\ref{tab:tt2L_nu_residuals}.
Relative uncertainties do not exceed 0.4\% on the Res. metric, do not exceed 0.75\% on the RMSE, and do not exceed 0.25\% for the $\Delta R$ mean. 
The EM($\Delta R$) method has the largest uncertainties, driven by a higher fraction of events failing reconstruction.

\begin{table*}[!t]
\centering
\caption{Event-by-event reconstruction accuracy metrics for four high-level observables in $t\bar{t}$(2L) production. 
For each observable, the row-normalized trace diagonal ratio and the effective resolution (Res.) central values are provided. 
The Res. of $m_{t\bar{t}}$ is quoted in units of GeV.
The maximum uncertainty on any metric does not exceed 0.5\%.
}
\label{tab:tt2L_highlevel}
\setlength{\tabcolsep}{10pt} 
\begin{tabular}{l cccccc cc}
\toprule
Technique & \multicolumn{2}{c}{$m_{t\bar{t}}$} & \multicolumn{2}{c}{$|y_t| - |y_{\bar{t}}|$} & \multicolumn{2}{c}{$\cos\theta_{K}^{+}$} & \multicolumn{2}{c}{$\cos\phi$} \\
\cmidrule(lr){2-3} \cmidrule(lr){4-5} \cmidrule(lr){6-7} \cmidrule(lr){8-9}
& Trace & Res. & Trace & Res. & Trace & Res. & Trace & Res. \\
\midrule

VyPER                & $0.520$ & $136$ & $0.619$ & $\textbf{0.812}$ & $0.579$ & $0.635$ & $0.593$ & $0.619$ \\
$\nu^2$-flows        & $0.518$ & $141$ & $\textbf{0.626}$ & $0.817$ & $0.576$ & $0.642$ & $0.588$ & $0.627$ \\
EM (VyPER)           & $\textbf{0.549}$ & $\textbf{128}$ & $0.593$ & $0.824$ & $\textbf{0.590}$ & $\textbf{0.609}$ & $\textbf{0.605}$ & $\textbf{0.593}$ \\
EM ($\Delta R$)       & $0.513$ & $150$ & $0.577$ & $0.891$ & $0.562$ & $0.673$ & $0.583$ & $0.640$ \\

\bottomrule
\end{tabular}
\end{table*}

\subsection{High-level observables}
Each neutrino is paired with a charged lepton, and
the edge assignment (as outlined in Appendix~\ref{apx:assign-strategy}) determine the \(b\)-jet associated to the lepton pair.
This defines a triad of final-state objects corresponding to three decay products of each top quark.
The sum of their respective four-momenta yields the kinematics of each parent quark, which are combined to reconstruct the \(t\bar{t}\) system. 
Given the similarity between the VyPER and SPANet assignment efficiencies, we focus henceforth exclusively on the VyPER model, omitting the SPANet results for brevity.

We assess the performance of each reconstruction method by examining several observables of interest to measurements in dileptonic $t\bar{t}$ production.
Figure~\ref{fig:tt2L_distributions_row} shows three such observables:
the invariant mass of the system, $m_{t\bar{t}}$;
the difference in absolute rapidity between the quarks, $|y_t| - |y_{\bar{t}}|$, used to measure asymmetries in production;
and the angular variable $\cos(\theta_k^+)$, sensitive to spin polarization and defined in Appendix~\ref{sec:def_angular_obs}.
An event from either EM implementation with a non-physical solution is omitted as before, and the histograms are normalized to unit area.
The ML-based techniques are seen to preserve the shape of the idealized reconstruction distribution better than the EM approach.

We evaluate the event-by-event fidelity of each reconstruction technique relative to the partonic truth.
Figure~\ref{fig:tt2L_migrations} show binned migration matrices for the $m_{t\bar{t}}$ and $\cos \phi$ variables, with the latter defined in Appendix~\ref{sec:appendix_assing}. 
While similar performance emerges across the reconstruction techniques for both observables, the exact fraction of events correctly retained on the main diagonal varies across the phase space.
Table~\ref{tab:tt2L_highlevel} summarizes the global performance characteristics, presented for the $m_{t\bar{t}}$, $\vert{}y_t\vert{} - \vert{}y_{\bar{t}}\vert{}$, $\cos\theta_k^+$, and $\cos\phi$ observables. Across three of the four observables, the EM approach with VyPER assignment yields the highest diagonal purity and the sharpest resolution. Furthermore, VyPER marginally outperforms $\nu^2$-flows in seven of the eight evaluated metrics, while the EM framework utilizing $\Delta R$ assignment consistently under-performs relative to the alternative techniques.

\subsection{Discussion}
When evaluating performance based on similarity to parton-level truth, 
fully ML-driven predictions outperform historic implementations of the EM method, implying that the precision of existing $t\bar{t}$ measurements could be enhanced through their adoption, particularly those that unfold to parton-level.
However, when studying the event-by-event comparison of high-level observables, the EM approach using VyPER-derived object assignments emerges as the optimal reconstruction strategy. 
We hypothesize that incorporating exact, on-shell mass constraints provides the EM with physical inductive bias that regularizes the reconstructed distributions. 
This finding demonstrates that analytically solving kinematic constraint equations remains a powerful approach provided the underlying combinatorial ambiguities --- such as the pairing of \(b\)-jets to leptons --- can be resolved with high accuracy in advance.
While comparisons utilizing top quarks defined within a fiducial particle-level phase space may yield a different performance ranking, such studies are left to future work.

\section{\label{sec:results_ww}Event reconstruction in the electroweak sector: $H \rightarrow W^\pm W^{\mp *}$ and VBS $W^\pm W^\pm$ production}
The precise study and characterization of the SM electroweak sector is one of the primary objectives of the LHC~\cite{ATLAS:2022Higgs10}. 
Processes like vector-boson fusion (VBF) Higgs production~\cite{CMS:2021vbf} and vector boson scattering~\cite{Covarelli:2021vbs} represent the frontier of our understanding of the mechanism of electroweak symmetry breaking, providing insight into the structure of the electroweak vacuum and the unitarization of scattering amplitudes via the Higgs mechanism at high energies~\cite{Lee:1977eg}.

Higgs boson production through VBF with a subsequent decay into a pair of $W$ bosons is a direct probe of the Higgs--vector-boson interaction~\cite{tumasyan2023measurements,ATLAS:2023hww_cp}, and event reconstruction will help constrain the CP properties of the coupling and map the spin structure of the Higgs boson~\cite{Anderson:2013afp,Fuchs:2020uoc}.
Measurements of this process could also leverage event reconstruction to improve the precision of Simplified Template Cross-Section (STXS) measurements~\cite{atlas25:hWWSTXS}, set limits on SMEFT operators and anomalous couplings~\cite{ATL-PHYS-PUB-2019-042,cms2024constraints}, and even observe quantum entanglement in the electroweak sector~\cite{barr2022testing}.

Vector-boson scattering is directly sensitive to the polarization structure of the scattering bosons, which drives the delicate interference between gauge and Higgs-mediated amplitudes.
Among all electroweak vector-boson scattering processes, the same-sign \(W^\pm W^\pm\) channel provides the cleanest experimental environment, making it the premier candidate for the study of boson polarization~\cite{aad2025evidence}, where event reconstruction will help extract these polarization fractions~\cite{panico2018diboson,ballestrero2020different}. 
Measurements of this process could also leverage these reconstructed kinematics to set new limits on anomalous quartic gauge couplings (aQGC) and dimension-eight SMEFT operators~\cite{degrande2023impact,aad2024measurement} and extend sensitivity to resonant and non-resonant new physics in the diboson invariant-mass tail~\cite{tumasyan2022search}.

Targeting these processes in leptonic channels provides clean experimental triggers and exceptional control over backgrounds, at the expense of a kinematically under-constrained final-state that has hampered the measurement of Higgs and electroweak boson properties.
Leptonic channels thus benefit most from full event reconstruction:
for both processes, we focus on the dileptonic channel where both $W$ bosons decay to a lepton-neutrino pair, of opposite electric charge in the $H \rightarrow WW^*$ process and of identical electric charge in the VBS $WW$ process.
We evaluate the performance of the VyPER and $\nu^2$-flows generative architectures to predict the kinematics of the neutrino paired with each charged lepton.

\subsection{Neutrino kinematics}
The comparisons of predicted neutrino kinematics to truth neutrino kinematics are given in Tables \ref{tab:hww_residuals} and \ref{tab:sswwjj_residuals} for the $H\rightarrow WW^*$ and VBS $WW$ processes, respectively.
The $H\rightarrow WW^*$ results show commensurate performance between both reconstruction techniques, while VyPER is seen to deliver more accurate neutrino kinematics in the VBS $WW$ process.

\begin{table}[tp]
\caption{Neutrino reconstruction accuracy for the $H\rightarrow WW^*$ process.
Metrics defined in Section \ref{subsec:metrics}.
Momentum metrics are quoted in GeV.
The relative uncertainty on any one metric does not exceed 0.3\%.
}
\label{tab:hww_residuals}
\begin{ruledtabular}
\begin{tabular}{lccccccc}
& $\Delta R$ & \multicolumn{2}{c}{$\Delta p_x$} & \multicolumn{2}{c}{$\Delta p_y$} & \multicolumn{2}{c}{$\Delta p_z$} \\
\cmidrule(lr){2-2} \cmidrule(lr){3-4} \cmidrule(lr){5-6} \cmidrule(lr){7-8}
Model & Mean & Res. & RMSE & Res. & RMSE & Res. & RMSE\\[1pt]\hline\\[-7pt]
VyPER          & \textbf{1.48}          & \textbf{39.7} & \textbf{24.0} & 39.7          & \textbf{24.1}          & \textbf{111} & \textbf{95.4} \\
$\nu^2$-flows  & \textbf{1.48} & \textbf{39.7} & 24.1          & \textbf{39.6} & \textbf{24.1} & 113          & 99.0          \\

\end{tabular}
\end{ruledtabular}
\end{table}

\begin{table}[tp]
\caption{Neutrino reconstruction accuracy for the VBS $WW$ process.
Metrics defined in Section~\ref{subsec:metrics}. 
Momentum metrics are quoted in GeV.
The relative uncertainty on any one metric does not exceed 0.5\%.
}
\label{tab:sswwjj_residuals}
\begin{ruledtabular}
\begin{tabular}{lccccccc}
& $\Delta R$ & \multicolumn{2}{c}{$\Delta p_x$} & \multicolumn{2}{c}{$\Delta p_y$} & \multicolumn{2}{c}{$\Delta p_z$} \\
\cmidrule(lr){2-2} \cmidrule(lr){3-4} \cmidrule(lr){5-6} \cmidrule(lr){7-8}
Model & Mean & Res. & RMSE & Res. & RMSE & Res. & RMSE\\[1pt]\hline\\[-7pt]
VyPER          & \textbf{1.55} & \textbf{82.0} & \textbf{57.6} & \textbf{80.9} & \textbf{56.7} & \textbf{214} & \textbf{193} \\
$\nu^2$-flows  & 1.70          & 93.6          & 67.5         & 93.9          & 67.5          & 246          & 229          \\
\end{tabular}
\end{ruledtabular}
\end{table}

\subsection{High-level observables}
Each neutrino is explicitly paired with a charged lepton prior to generating the neutrino kinematics.
The \(W\) bosons are then reconstructed by summing the four-momenta of the pair.
For the \(H\rightarrow WW^*\) channel, the four-momenta of the two \(W\) bosons are further summed to reconstruct the Higgs boson.
In the idealized baseline, the true neutrino kinematics are used instead of the model predictions.

\begin{figure*}
    \centering
    \includegraphics[width=\textwidth]{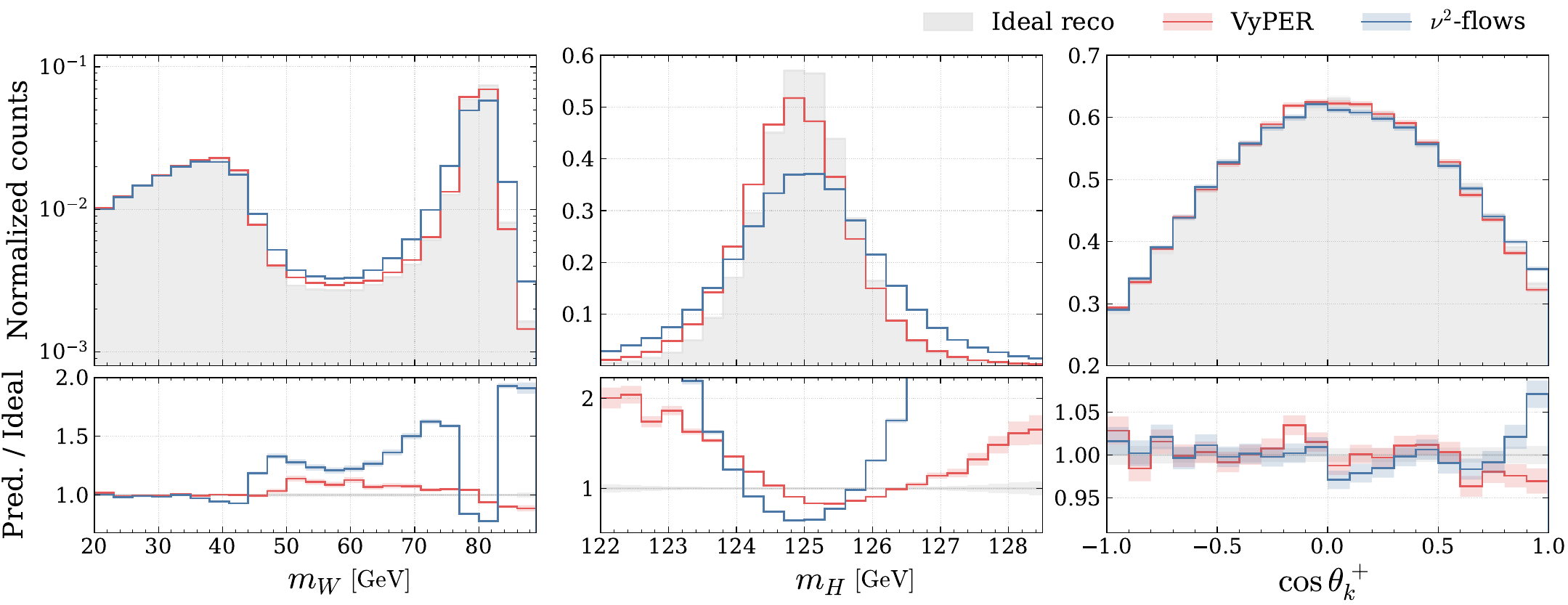}
    \caption{\label{fig:HWW_distributions} 
    Kinematic distributions of high-level observables in $H\rightarrow WW^*$ production.
    The $W$ bosons are reconstructed by combining each lepton's four-momentum with either the true neutrino kinematics (for the ``Ideal reco'' baseline) or the model predictions from VyPER and $\nu^2$-flows.
    The Higgs boson four-momentum is given by the sum of the two $W$ bosons' four-momenta.
    Shaded bands illustrate the statistical uncertainty in each bin.}
\end{figure*}

For each process, we select an individual set of observables constructed from the reconstructed \(W\) boson kinematics; these are chosen either to evaluate reconstruction performance or because they represent observables relevant for downstream physics measurements.
Figure~\ref{fig:HWW_distributions} shows such high-level observables for $H \rightarrow WW^*$ and demonstrates that VyPER recovers the double-peak structure of W boson invariant mass distribution, $m_W$ (corresponding to the on-shell and off-shell $W$ bosons) better than $\nu^2$-flows, and similarly more closely matches the idealized shape of the Higgs boson invariant mass, $m_H$.
Figure~\ref{fig:HWW_distributions} also shows the cosine of the $W^+$ helicity angle in the Higgs boson reference frame (defined in Appendix~\ref{sec:def_angular_obs}), where reconstruction performance is found to be similar with higher deviations towards positive unity. 
Table~\ref{tab:hww_metrics} quotes the event-by-event similarity metrics for $m_H$ and $\cos\theta_{K}^{+}$, with $m_W$ replaced by the transverse momentum of the Higgs boson, $p_T^H$. 
We observe that VyPER reconstruction marginally out-performs $\nu^2$-flows.

\begin{table}[h!]
\caption{Event-by-event similarity metrics for $H\rightarrow WW^*$ production.
Resolution (Res.) of $m_{H}$ and $p_{\mathrm{T}}^{H}$ are quoted in GeV.
All relative uncertainties on the trace fall below 0.2\% and all relative uncertainties on the Res. fall below 0.3\%.
}
\label{tab:hww_metrics}
\begin{ruledtabular}
\begin{tabular}{lcccccc}
                & \multicolumn{2}{c}{$m_{H}$} & \multicolumn{2}{c}{$p_{\mathrm{T}}^{H}$} & \multicolumn{2}{c}{$\cos\theta_{K}^{+}$} \\[1pt]\cline{2-3}\cline{4-5}\cline{6-7}\\[-9pt]
Models          & Trace & Res. & Trace & Res. & Trace & Res.   \\[1pt]\hline\\[-7pt]
VyPER           & \textbf{0.270} & \textbf{2.22} & \textbf{0.675} & \textbf{29.0} & \textbf{0.423} & \textbf{0.992} \\[1pt]
$\nu^2$-flows             & 0.266          & 2.91          & 0.668          & 30.1          & 0.419          & 1.00 
\end{tabular}
\end{ruledtabular}
\end{table}

\begin{figure*}
    \centering
    \includegraphics[width=\textwidth]{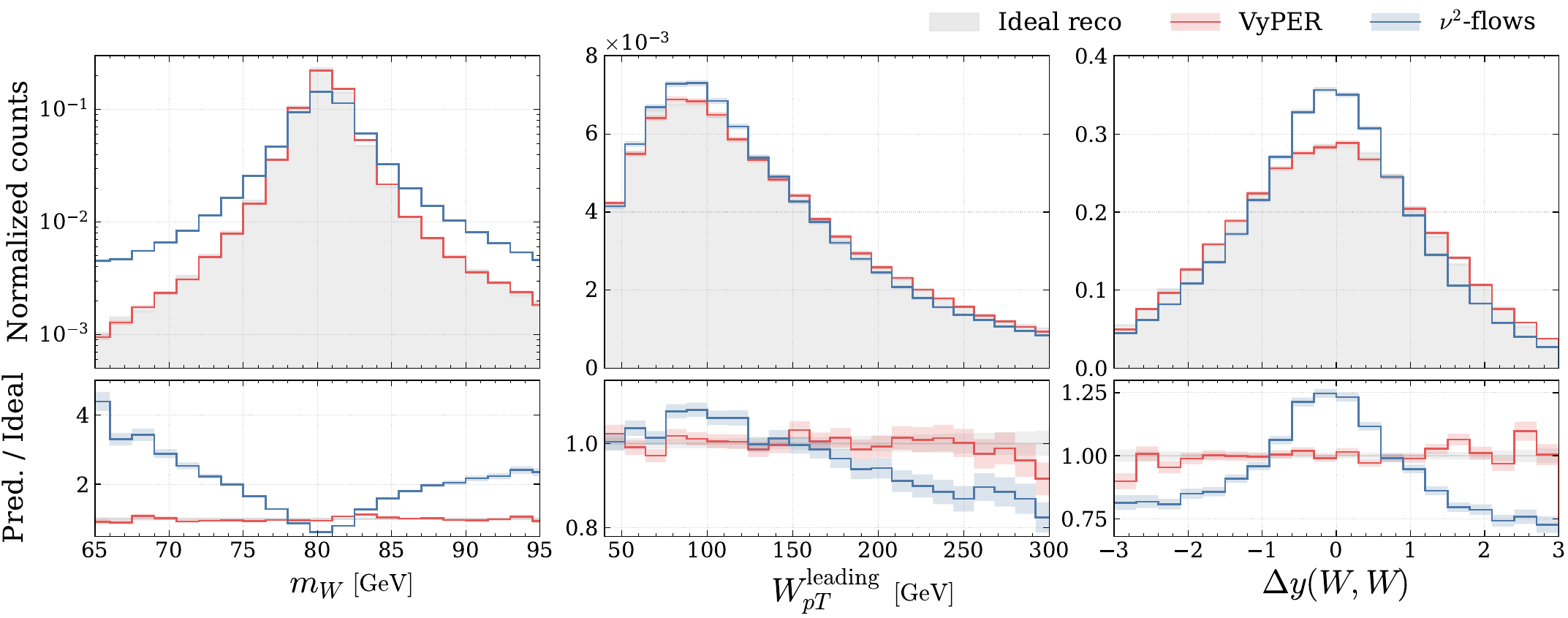}
    \caption{ \label{fig:sWW_distributions} 
    Kinematic distributions of high-level observables in VBS $WW$ production.
    The $W$ bosons are reconstructed by combining each lepton's four-momentum with either the true neutrino kinematics (for the ``Ideal reco'' baseline) or the model predictions from VyPER and $\nu^2$-flows.
    Shaded bands indicate the statistical uncertainty in each bin.}
\end{figure*}

VyPER's superior neutrino reconstruction accuracy in VBS $WW$ translates into superior replication of the idealized reconstructed distribution shapes in Figure~\ref{fig:sWW_distributions}. 
This figure evaluates the \(W\) boson mass, \(m_{W}\), as a direct benchmark for the resolution of the reconstruction technique. 
It also shows the leading \(W\) boson transverse momentum, \(W_{pT}^{\text{leading}}\), and the rapidity difference between the bosons, which serve as key observables for studying electroweak symmetry breaking: \(W_{pT}^{\text{leading}}\) probes the high-energy tail sensitive to aQGCs, while the rapidity separation isolates the characteristic scattering topology of the VBS process.
Table~\ref{tab:ssww_high_sim} summarizes the event-by-event similarity between the predictions and the idealized reconstruction. 
Here, the invariant mass of the diboson system, \(m_{WW}\), is evaluated instead of the individual \(W\) boson mass, as \(m_{WW}\) is thought to be a more relevant observable for physics interpretations and cross-section measurements.
VyPER realises a superior event reconstruction across all metrics.

\begin{table}[h!]
\caption{Event-by-event similarity metrics for VBS $WW$ production.
Metrics defined in Section~\ref{subsec:metrics}. 
Resolution (Res.) of $m_{WW}$ and $W_{pT}^{\mathrm{leading}}$ are quoted in GeV.
All relative uncertainties on the trace fall below 0.25\% and all relative uncertainties on the Res. fall below 0.5\%. }
\label{tab:ssww_high_sim}
\begin{ruledtabular}
\begin{tabular}{lcccccc}
                & \multicolumn{2}{c}{$m_{WW}$} & \multicolumn{2}{c}{$W_{pT}^{\mathrm{leading}}$} & \multicolumn{2}{c}{$\Delta y(W,W)$} \\[1pt]\cline{2-3}\cline{4-5}\cline{6-7}\\[-9pt]
Model         & Trace & Res. & Trace & Res. & Trace & Res.   \\[1pt]\hline\\[-7pt]

VyPER           & \textbf{0.420} & \textbf{180} & \textbf{0.552} & \textbf{81.8} & \textbf{0.630} & \textbf{1.47} \\[1pt]
$\nu^2$-flows             & 0.412          & 186          & 0.517          & 90.0          & 0.544          & 1.76 
\end{tabular}
\end{ruledtabular}
\end{table}

\subsection{Discussion}
We show that generative techniques accurately model neutrino kinematics in electroweak processes. 
This enables the high-fidelity reconstruction of high-level \(W\) boson and Higgs boson observables, including both kinematic and angular distributions essential for probing boson couplings and polarizations. 
In particular, VyPER demonstrates a superior ability to reproduce reconstructed mass spectra in both $H \rightarrow WW^*$ and VBS \(WW\) processes, and yields greater similarity to truth on an event-by-event basis for all studied observables. 
The application of these techniques establishes a new paradigm in multi-lepton electroweak measurements:
one where high-level observables are fully reconstructed and used to probe the electroweak sector with greater scrutiny.

\section{\label{sec:ttW} Event reconstruction in $t\bar{t}W^\pm$ production}
The production of top-antitop quark pairs in association with a $W$ boson offers invaluable insight into the couplings between electroweak bosons and the top quark sector.
This rare scattering process has been measured inclusively and differentially at the LHC~\cite{ATLAS:2015ttWttZ8TeV,CMS:2015ttWttZ8TeV,ATLAS:2024ttW13TeV,CMS:2026ttWdiff13TeV}, constitutes a key background in the measurements of other rare processes including $t\bar{t}H$ production~\cite{ATLAS:2025tth}, and is an important component of global SMEFT fits in the top quark sector~\cite{vonBuddenbrock:2020ter,CMS:2023btv,ATLAS:2026ttwjsmeft}.
Event reconstruction in this process would enable unfolding to the parton level, potentially shedding light on the origin of the persistent discrepancies between measured inclusive rates and the most precise fixed-order predictions~\cite{Denner:2021gqi,Buonocuore:2023vdg}. 
Furthermore, a broader suite of properties can be explored: full reconstruction facilitates the measurement of partonic charge asymmetries as well as variables sensitive to the polarization of the $W$ boson itself~\cite{Maltoni:2014zpa,CMS:2026ttWdiff13TeV}.
With its complex and diverse final state, $t\bar{t}W^\pm$ production serves as a rich playground for benchmarking event reconstruction techniques, and allows us to showcase VyPER's edge reconstruction, hyperedge reconstruction and neutrino prediction capabilities in a single topology.
We study the two-lepton final state where both leptons have the same electric charge: the associated prompt $W$ boson must always decay leptonically, with one top quark also decaying leptonically and the other hadronically.

\subsection{Assignment}
The assignment task looks to assign three jets to the hadronic top, and to pair a lepton with a jet to define the leptonic top.
The lepton not paired is then identified as coming from the associated $W$ boson decay.
VyPER and SPANet are trained to perform this assignment.
The assignment procedure for VyPER is detailed in Appendix.~\ref{apx:assign-strategy}.
SPANet is observed to produce the unphysical combination of two leptons assigned to the same top quark in a small number of cases:
these events are discarded from all comparisons for all networks.
The performance of both networks is tabulated in Table.~\ref{tab:ttw-efficiency}, where it is evident that VyPER achieves a higher reconstruction efficiency for all components of the system. 

\begin{table}[!ht]
\centering
\caption{Inclusive assignment efficiency of VyPER and SPANet in 2LSS $t\bar{t}W$ events for the $W$ boson from the hadronic top quark ($\varepsilon_{W_{\text{had}}}$), the hadronic top quark ($\varepsilon_{t_{\text{had}}}$), the leptonic top quark ($\varepsilon_{t_{\text{lep}}}$), and the associated $W$ boson ($\varepsilon_{W_{\text{assoc}}}$). The event efficiency $\varepsilon_{t\bar{t}W}$ requires all four objects to be correctly reconstructed simultaneously. All values are reported in percentages (\%).
The absolute uncertainty on any one value does not exceed 0.1\%.}
\label{tab:ttw-efficiency}
\begin{ruledtabular}
\begin{tabular}{lccccc}
Model & $\varepsilon_{W_{\text{had}}}$ [\%] & $\varepsilon_{t_{\text{had}}}$ [\%] & $\varepsilon_{t_{\text{lep}}}$ [\%] & $\varepsilon_{W_{\text{assoc}}}$ [\%] & $\varepsilon_{t\bar{t}W}$ [\%] \\[4pt] \hline
\rule{0pt}{14pt}VyPER  & $\mathbf{89.7}$ & $\mathbf{83.5}$ & $\mathbf{73.6}$ & $\mathbf{80.4}$ & $\mathbf{68.8}$ \\
SPANet & 86.4            & 80.3            & 70.1            & 78.2            & 64.6            \\
\end{tabular}
\end{ruledtabular}
\end{table}

\subsection{Neutrino reconstruction}

Table~\ref{tab:ttw_residuals} presents the neutrino reconstruction accuracy, where we observe that VyPER outperforms $\nu^2$-flows in all metrics.

\begin{table}[tp]
\caption{Neutrino reconstruction accuracy for the same-sign $t\bar{t}W$ process.
Metrics defined in Section \ref{subsec:metrics}.
Momentum metrics are quoted in units of GeV.
The relative uncertainty on any metric does not exceed 0.35\%.}
\label{tab:ttw_residuals}
\begin{ruledtabular}
\begin{tabular}{lccccccc}
& $\Delta R$ & \multicolumn{2}{c}{$\Delta p_x$} & \multicolumn{2}{c}{$\Delta p_y$} & \multicolumn{2}{c}{$\Delta p_z$} \\
\cmidrule(lr){2-2} \cmidrule(lr){3-4} \cmidrule(lr){5-6} \cmidrule(lr){7-8}
Model & Mean & Res. & RMSE & Res. & RMSE & Res. & RMSE\\[1pt]\hline\\[-7pt]
VyPER          & \textbf{1.46} & \textbf{75.7} & \textbf{48.6} & \textbf{75.0} & \textbf{48.5} & \textbf{188} & \textbf{174} \\
$\nu^2$-flows  & 1.59          & 87.1          & 54.7          & 87.6          & 54.6          & 212          & 198          \\
\end{tabular}
\end{ruledtabular}
\end{table}

\subsection{High-level observables}
Prior to kinematic reconstruction, neutrinos are paired with charged leptons. 
Following the strategy outlined in Appendix~\ref{apx:assign-strategy}, VyPER assigns the remaining jets and lepton-neutrino pairs to define the constituents of the hadronic top, leptonic top, and associated \(W\) boson.
The kinematics of each individual parent particle and of the overall system are then reconstructed analogously to the previous channels. 
We compare high-level observables constructed using VyPER to those obtained by combining \nflows neutrino predictions with either the VyPER or SPANet assignment configurations.

\begin{figure*}
    \centering
    \includegraphics[width=\textwidth]{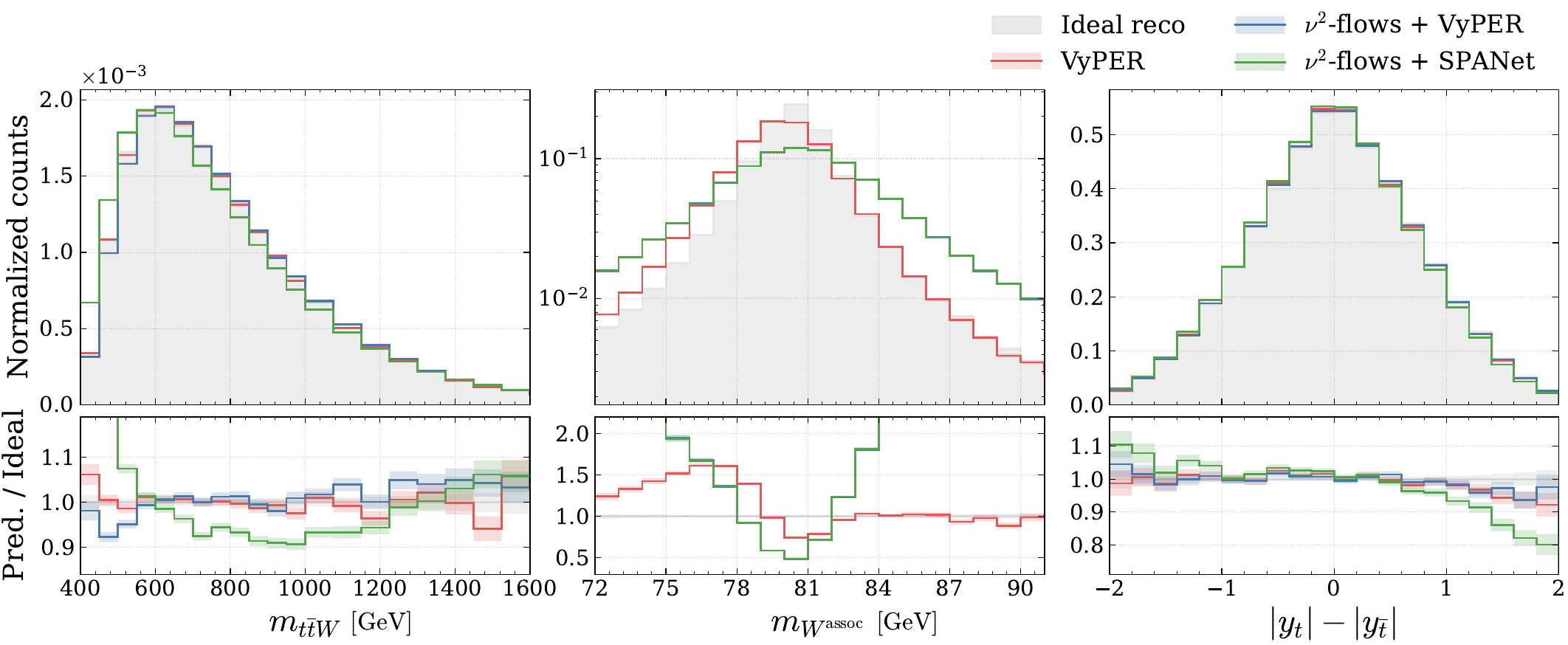}
    \caption{\label{fig:ttW_distributions_row} Kinematic distributions of high-level observables in $t\bar{t}W$ production.
    Neutrinos predictions are produced by both VyPER and $\nu^2$-flows.
    Charged lepton-neutrino pairs are assigned along with jets by both VyPER and SPANet to define both top quarks and the associated $W$ boson.
    The VyPER assignment order is described in Appendix~\ref{apx:assign-strategy}.}
\end{figure*}

Distributions of the overall invariant mass of the system $m_{t\bar{t}W}$, the mass of the associated $W$ boson $m_{W^{\text{assoc}}}$, and the difference in absolute rapidity between the top and antitop quarks, $|y_t| - |y_{\bar{t}}| $, are shown in Figure~\ref{fig:ttW_distributions_row}.
The distributions built using VyPER assignment show greater similarity to the idealized reconstruction, and the associated $W$ boson mass is only captured with high fidelity using the VyPER-predicted neutrinos.
Differences in $|y_t| - |y_{\bar{t}}|$, the critical observable for charge asymmetry measurements, are less pronounced.

\begin{table*}[tbp]
\centering
\caption{Reconstruction performance metrics comparison for the $t\bar{t}W$ process across different observables and techniques. 
The combination used for neutrino prediction and assignment are given in the first two columns respectively.
The $m_{t\bar{t}W}$ and $p_{T}^{W}$ variables are in units of GeV.
The relative uncertainty on the trace does not exceed 0.25\% and the relative uncertainty on the Res. does not exceed 0.35\%.}
\label{tab:ttw_high} 
\footnotesize 
\setlength{\tabcolsep}{6.5pt} 
\begin{tabular}{ll cccccc cc}
\toprule
\multicolumn{2}{c}{Technique} & \multicolumn{2}{c}{$m_{t\bar{t}W}$} & \multicolumn{2}{c}{$p_{T}^{W}$} & \multicolumn{2}{c}{$|y_t| - |y_{\bar{t}}|$} & \multicolumn{2}{c}{$\cos\theta^{*}$} \\
\cmidrule(lr){1-2} \cmidrule(lr){3-4} \cmidrule(lr){5-6} \cmidrule(lr){7-8} \cmidrule(lr){9-10}
Neutrino Reco. & Assignment & Trace & Res. & Trace & Res. & Trace & Res. & Trace & Res. \\
\midrule
VyPER          & VyPER      & $\textbf{0.672}$ & $\textbf{105}$ & $\textbf{0.622}$ & $\textbf{43.5}$ & $\textbf{0.717}$ & $\textbf{0.300}$ & $\textbf{0.660}$ & $\textbf{0.221}$ \\
$\nu^2$-flows  & VyPER      & $0.650$          & $114$          & $0.590$          & $48.0$          & $0.702$          & $0.326$          & $0.646$          & $0.239$          \\
$\nu^2$-flows  & SPANet     & $0.601$          & $142$          & $0.586$          & $48.7$          & $0.619$          & $0.440$          & $0.609$          & $0.281$          \\

\bottomrule
\end{tabular}
\end{table*}

In Table~\ref{tab:ttw_high}, the comparison of each prediction to the partonic truth is summarized for four observables: $m_{t\bar{t}W}$ and  $|y_t| - |y_{\bar{t}}| $, as well as the transverse momentum of the associated $W$ boson $p_T^W$, and $\cos \theta^*$, the scattering angle of the top quark in the $t\bar{t}W$ center-of-mass frame, defined in Appendix~\ref{sec:def_angular_obs}.
VyPER exhibits the largest trace and smallest Res. for all four observables.

\subsection{Discussion}
Reconstructing the complete \(t\bar{t}W\) final state requires a combination of neutrino kinematic prediction and combinatorial particle assignments.
VyPER integrates these distinct challenges into a single, unified learning objective, outperforming SPANet in all assignment metrics and \(\nu ^{2}\)-flows in neutrino prediction accuracy. 
This approach yields superior modeling of idealized reconstruction distributions of high-level kinematic and angular observables, alongside tighter resolution relative to parton-level observables. 
This illustrates how event reconstruction can be successfully applied to high-multiplicity processes with more than two parent particles, presenting new opportunities to probe SM properties in multi-top-quark and rare Higgs boson production, or search for new physics in heavy supersymmetric decay cascades.

\section{\label{sec:conclusion}Conclusions}
Event reconstruction remains a critical task in the analysis of particle collider data, invaluable for producing unfolded top quark kinematic spectra and precisely measuring electroweak boson properties.
We factorized event reconstruction into the separate tasks of assigning measured jets and charged leptons to parent particles, and predicting unmeasured neutrino kinematics, remarking that existing reconstruction tools generally solve only one of these problems.
VyPER was presented as a framework designed to solve both tasks simultaneously, providing comprehensive event reconstruction for arbitrary SM physics processes.

The performance of several event reconstruction tools were compared to VyPER in four different physics processes.
In dileptonic \(t\bar{t}\) production, utilizing VyPER-derived \(b\)-jet–lepton pairings as input for the analytical Ellipse Method yielded the highest resolution between the reconstructed and parton-level observables, demonstrating that a hybrid machine-learning and analytical framework offers a powerful avenue for future $t\bar{t}$ measurements.
In electroweak \(H \rightarrow WW^*\) and VBS \(WW\) scattering processes, generative machine learning techniques successfully captured the correlations between measured final states and neutrino kinematics, enabling the accurate reconstruction of these inherently under-constrained bosonic systems, opening new avenues for measuring specific boson properties in multi-lepton channels.
In $t\bar{t}W$ production, VyPER paired superior assignment efficiency with accurate neutrino predictions to reconstruct observables with high resolution.
Reconstruction techniques that can perform accurate assignment and neutrino prediction will pave the way for precision measurements of top-quark properties in these rare production modes using the expanded LHC Run 3 datasets.

Beyond introducing the VyPER framework, this work demonstrates that unifying efficient assignment with precise neutrino kinematic prediction enables the accurate recovery of short-lived parent particles in arbitrary final states with a single tool.
Ultimately, individual analyses are advised to evaluate the performance of reconstruction techniques based on their specific downstream sensitivity and final measurement goals.
We anticipate full event reconstruction having utility in channels beyond those explored here, and await published experimental measurements which make use of comprehensive reconstruction models like VyPER.
The challenge of event reconstruction will continue to drive novel ML reconstruction development, which in turn will help deliver a more precise scrutiny of the SM than ever before using data collected at the High-Luminosity LHC.

\begin{acknowledgments}
Y.P. and E.S. are supported by UK Research and Innovation [grant number EP/Z533865/1]. The project was selected by the ERC, and funded by UKRI.
\end{acknowledgments}

\appendix

\section{Additional information on VyPER}
VyPER is an open source \textsc{Python} project created with the \textsc{Pytorch}~\cite{NEURIPS2019_9015} and \textsc{Pytorch-Geometric}~\cite{Fey_Lenssen_2019} libraries.
Its training and inference frameworks are built using \textsc{Pytorch-Lightning}~\cite{falcon2020pytorchlightning}.
The code is freely available on GitHub~\footnote{\url{https://github.com/tzuhanchang/VyPER}}.

In this appendix, we describe the network inputs and hyperparameters used in the presented studies, as well as the assignment strategy used to reconstruct candidates from the edge and hyperedge probabilities given by VyPER.

\subsection{\label{apx:input} Network input}
The initial state of the graph, that is, at message-passing step $s=0$, is defined by the kinematics of the final-state objects, their pairwise relations, and global event information:
\begin{eqnarray}
    \label{eq:node_input}
    \mathbf{x}^{(0)}_i &=& \left( {p_T}_i, \eta_i, \phi_i, E_i, Q_i, b\textrm{-tag}_i, \mathrm{ID} \right),\\
    \label{eq:edge_input}
    \mathbf{e}^{(0)}_{ij} &=& ( \Delta\eta_{ij}, \Delta\phi_{ij}, \Delta R_{ij}, M_{ij}, k_{T,{ij}}, z_{ij} ),  \\
    \label{eq:glob_input}
    \mathbf{u}^{(0)} &=& ( N_\mathrm{jets}, N_\mathrm{b\textrm{-tagged}}, N_\mathrm{e}, N_\mathrm{\mu}, E_T^{\rm miss}, \nonumber \\
    & &\hfill \phi_{E_T^{\rm miss}}, \cos{\phi_{E_T^{\rm miss}}}, \sin{\phi_{E_T^{\rm miss}}}  ),
\end{eqnarray}
where $i,j\in V_\mathrm{exp}$ represents the $i$-th and $j$-th final state.

The node inputs listed in Eq.~\ref{eq:node_input} pertain to the transverse momentum, pseudorapidity, azimuth angle, energy, charge, $b$-tagging status, and type of the object, respectively.
Charge and $b$-tagging status is not applicable to all final-state objects: for jets, we set $Q=0$, and for leptons, we set $b\textrm{-tag}=0$.
For object type, $\mathrm{ID}=0$ is used for jets; $1$ and $2$ are used for electrons and muons, respectively.

The relations between each $ij$ final-state pair are computed and used as edge inputs.
The features comprise their angular separation ($\Delta\eta_{ij}$, $\Delta\phi_{ij}$, and $\Delta R_{ij}$), their combined invariant mass ($M_{ij}$),  as well as $k_{T,ij}= \min(p_{T,i}, p_{T,j})\Delta R$ and $z_{ij} = \min(p_{T,i}, p_{T,j}) / (p_{T,i} + p_{T,j})$.
The global input includes the multiplicity of jets, $b$-tagged jets, electrons and muons, as well the missing transverse momentum and its azimuth direction, along with its cosine and sine, as listed in Eq.~\ref{eq:glob_input}.

\subsection{\label{apx:hparam} Network configurations}
A set of hyperparameters used in our studies is shown in Table~\ref{tab:hyperparameters}.
In the $t\bar{t}$ experiment, both edge classification and neutrino diffusion components are enabled, yielding a model size of 3.6 million trainable parameters, whereas the two neutrino reconstruction-only tests ($H\rightarrow WW^*$ and VBS same-sign $WW$) use a smaller model of 3.2 million parameters.
With all three components active, the model used for the $t\bar{t}W$ reconstruction has a 5 million trainable parameters.

\begin{table}[h!]
\renewcommand{\arraystretch}{1.25}
\caption{\label{tab:hyperparameters}
A list of hyperparameters used to setup VyPER for the presented study.
}
\begin{ruledtabular}
\begin{tabular}{clccc}
& Name & Expression & Variable & Value \\
\colrule
1 & \makecell[l]{Number of\\MPNN steps} & $s=1,\dots, S$ & $S$ & 3 \\
2 & \makecell[l]{Message\\dimensionality} & $\mathbf{x}_i^{(s)},\mathbf{e}_{ij}^{(s)},\mathbf{u}^{(s)}\in\mathbb{R}^D$ & $D$ & 128 \\
3 & \makecell[l]{Hyperedge\\dimensionality} & $\mathbf{v}_m\in\mathbb{R}^L$ & $L$ & 128 \\
4 & \makecell[l]{Context\\dimensionality} & $\mathbf{x}^*_k, \mathbf{T}_\theta(t)\in\mathbb{R}^W$ & $W$ & 128 \\
5 & \makecell[l]{Number of\\diffusion steps} & $t=1,\dots, T$ & $T$ & 200 \\
6 & \makecell[l]{Number of\\DiT modules} & --- & --- & 4 \\
7 & \makecell[l]{Number of\\attention heads} & --- & --- & 8 \\
8 & \makecell[l]{Hyperedge\\loss scale} & $\alpha$ & $\alpha$ & 0.5 \\
9 & \makecell[l]{Diffusion\\loss scale} & $\eta$ & $\eta$ & 0.8~\footnote{Set to 0.6 for the dileptonic $t\bar{t}$ study.} \\[6pt]
10& Learning rate & --- & --- & 0.00025 \\[4pt]
11& Dropout & --- & --- & 0.001 \\[4pt]
12& Batch size & --- & --- & 4096 \\
\end{tabular}
\end{ruledtabular}
\end{table}

The learning rate follows a schedule based on the recorded validation loss: it is reduced by a factor of 0.8 whenever the validation loss has not improved for 10 epochs.
An early stopping mechanism is employed to terminate the training after 50 non-improving epochs.
Only the network state that produces the lowest validation loss is saved.
All experiments were run on an NVIDIA DGX Spark GB10 (128GB) with CUDA 13.0.

\subsection{\label{apx:assign-strategy}Assignment strategy}
VyPER outputs a score for each edge and hyperedge, representing the probability that the given structure contains the correct final-state products of a parent particle decay. These probabilities are used to determine the final assignment with channel-specific selection criteria.

For $t\bar{t}$~(2L), we choose the two highest-scoring edges, each of which must have exactly one lepton as an endpoint and  share no common nodes.

For $t\bar{t}W$ (2LSS), VyPER predicts two scores for each edge, corresponding to the leptonic-top edge class (connecting the $b$-jet and lepton from the leptonic top) and the hadronic-$W$ class (connecting the two jets from the hadronic $W$ boson).
All 3-node hyperedges are simultaneously classified to give a probability that they constitute the three jets from the hadronic top.
We prioritize the reconstruction of the leptonic top quark, choosing whichever edge has the highest leptonic-top class score.
We then compute a total hadronic-top score by summing the three individual hadronic-$W$ scores within each triplet of jets, and adding the corresponding hyperedge score. 
Based on this combined score, we select the highest scoring candidate that shares no nodes with the pre-selected leptonic top quark edge.

\section{\label{sec:appendix_nu2flows_spanet} Implementation of SPANet and $\nu^2$-flows}
In the presented studies, VyPER is benchmarked against two ML techniques: SPANet~\cite{spanet2021,spanet2022,spanet2024} and $\nu^2$-flows~\cite{nu2flows2024}, whose training setups are described in this section.
We use the latest SPANet release, v2.3 (available at~\footnote{\url{https://github.com/Alexanders101/SPANet/releases/tag/v2.3}}), configured using the setup recommended by its authors for all-hadronic $t\bar{t}$, adapted for the specific processes considered here.

The public version of $\nu^2$-flows (available at~\footnote{\url{https://github.com/rodem-hep/nu2flows/tree/for\_public}}) is employed in our studies.
We follow the authors' recommended setup for dileptonic $t\bar{t}$, with minor adjustments to match our specific processes.
$\nu^2$-flows outputs a neutrino and an anti-neutrino, which can be unambiguously assigned to the correct lepton for events with two oppositely charged leptons.
For events with a same-sign lepton pair, $\nu^2$-flows requires a minor modification, in which case each neutrino is implicitly associated to a lepton in the same way as in VyPER.

\section{\label{sec:appendix_assing} Assignment models in the literature}
Supervised ML-based assignment algorithms embed event information into high-dimensional latent spaces through a sequence of learnable transformations, and then utilize specific classification heads to solve the assignment.
The assignment strategy varies by model.
The SPANet \cite{spanet2021, spanet2022, spanet2024} framework optimizes a global categorical cross-entropy loss over all possible final-state combinations, forcing all decay configurations to compete simultaneously.
The SAJA \cite{lee2024zero, heo2025improving} model classifies each final-state object individually, minimizing a cross-entropy loss across a set of target parton labels.
The HyPER model \cite{hyper2024} casts all object combinations as independent hyperedges and classifies these using cross-entropy loss.
The Topograph \cite{topograph2023} and TIGER \cite{soybelman2026topology} models leverage graph structures to recursively classify binary edges linking observed objects to intermediate parent nodes.
The 0-lepton $t\bar{t}$ channel has been a focus of model comparison: recent studies have shown that leading reconstruction efficiencies can be achieved with smaller graph-based models compared to larger transformers \cite{hyper2024}, that dropping candidate parent particles with low reconstruction scores can yield improved reconstruction purities \cite{soybelman2026topology}.
The latest work in this area applies the generative ML paradigm in the discrete case \cite{hermansen2026pairton} to iteratively solve the assignment problem.

\section{\label{sec:ttbar_alt} Review of alternative $t\bar{t}$(2L) reconstruction methods}
\subsection{Sonnenschein method}
The Sonnenschein method casts the constrained kinematic equations of the $t\bar{t}$ decay system as a pair of quadratics in the neutrino and anti-neutrino longitudinal momentum \cite{sonnenschein2005algebraic}.
The system is reformulated as a quartic polynomial in one component of neutrino momenta and solved through the method of resultants.
As stated in Section~\ref{subsec:tt2L_ellipse}, the Sonnenschein method was not tested as it was found historically to perform worse than the EM.

\subsection{NeutrinoWeighter method}
The NeutrinoWeighter (NW) \cite{NW_D0_1999} method does not explicitly solve kinematic equations of constraint, instead repeatedly sampling a simulated distribution of neutrino pseudorapidity $\eta^{\nu}$ to give an ensemble of solution hypotheses per event.
Each hypothesis is assigned a weight that is Gaussian in the difference between the predicted combined neutrino transverse momentum and the MET, with the highest weighted hypothesis selected as neutrino candidates for that event.

A baseline implementation of the NW algorithm is not provided in the literature; further, it is known to be extremely computationally-expensive, particularly when implementing a fine-grained $\eta$ sampling.
To this end, we implement the NW algorithm in \textsc{PyTORCH} for parallel execution on graphical processing units (GPUs).
This implementation computes neutrino solutions for batches of events with a batch size of 64, in 50 $\eta^{\nu}$ steps.
The sampling of top quark and $W$ boson masses is identical to the EM case above.
Our implementation is available at \href{https://github.com/tzuhanchang/TensorNW}{https://github.com/tzuhanchang/TensorNW}.

As stated in Section~\ref{subsec:tt2L_ellipse}, we applied this implementation of the NW method but chose not to present the results as they were inferior to the EM.

\section{\label{sec:def_angular_obs}Construction of angular observables}
Angular observables defined in reference frames other than the laboratory frame offer a stern test to reconstruction algorithms, as the combination of Lorentz boosts and vector operations place a heavy demand on attaining accurate assignment and neutrino kinematics.

In the $t\bar{t}$ and $H\rightarrow WW^*$ studies, the observable $\cos \theta^+_k$ is the cosine of the ``helicity angle'', defined in the helicity basis with definition given in~\cite{bernreuther2015set}. 
In $t\bar{t}$(2L) the observable is constructed by first Lorentz boosting the reconstructed top quarks and the leptons into the $t\bar{t}$ center-of-mass frame, and then further boosting the leptons into their individual parent top quarks' frames as defined in the $t\bar{t}$ center-of-mass frame.
The variable is then given by the scalar product of the positive lepton's spatial direction and the beam line.
The $H\rightarrow WW^*$ process is completely analogous given the substitution of the $W$ bosons for the top quarks.
The $\cos \phi$ observable is defined as scalar product between the two leptons defined as above in their parent reference frame. 
This observable is only studied in $t\bar{t}$(2L).

In the $t\bar{t}W$ study, we examine the cosine of the top quark production angle, $\cos \theta^*$, which is simply the cosine of the angle between the top quark -- as defined in the $t\bar{t}$ center-of-mass frame -- and the beam line.

\section{\label{sec:soft_data_avail}Data and software availability}
VyPER code is available on GitHub at \url{https://github.com/tzuhanchang/VyPER/}.
Detailed instructions on how to perform training and inference are provided in the README.md.

Training, validation and testing datasets are made available for all four physics processes on Zenodo  \cite{mao_2026_22308461}(\url{https://doi.org/10.5281/zenodo.22308461}) and HuggingFace (\url{https://huggingface.co/datasets/tzuhanchang/VyPER}).
The records describes the detector-level and generator-level information saved in ROOT format \cite{antcheva2009root}.


\bibliography{bibliography}

\end{document}